\documentclass[reprint, 
nofootinbib,
amsmath,amssymb,
aps,prd
]{revtex4}

\usepackage{graphicx}
\usepackage{bm}
\usepackage{booktabs}
\usepackage{amsmath}
\usepackage{mathrsfs}
\usepackage{subfigure}
\usepackage{caption}
\usepackage[hidelinks]{hyperref}
\hypersetup{
    colorlinks=true,
    linkcolor=blue,
    filecolor=gray,      
    urlcolor=blue,
    citecolor=blue,
}
\allowdisplaybreaks[1]

\begin{document}


\title{Post-Newtonian dynamics of compact binaries with mass transfer}

\author{Zi-Han Zhang$^{1,2}$}
 \email{zhangzihan242@mails.ucas.ac.cn}

\author{Tan Liu$^{3,4}$}
 \email{lewton@mail.ustc.edu.cn}

\author{Zong-Kuan Guo$^{3,4,5}$}
\email{guozk@itp.ac.cn}

\affiliation{$^{1}$International Centre for Theoretical Physics Asia-Pacific, University of Chinese Academy of Sciences, 100190 Beijing, China}
\affiliation{$^{2}$Taiji Laboratory for Gravitational Wave Universe, University of Chinese Academy of Sciences, 100049 Beijing, China}

\affiliation{$^{3}$School of Fundamental Physics and Mathematical Sciences, Hangzhou Institute for Advanced Study, University of Chinese Academy of Sciences, Hangzhou 310024, China}

\affiliation{$^{4}$School of Physical Sciences, University of Chinese Academy of Sciences, Beijing 100049, China }
\address{$^{5}$Institute of Theoretical Physics, Chinese Academy of Sciences, Beijing 100190, China }


\date{\today}

 \begin{abstract}
Taking into account the mass transfer effect, we derive the equations of motion of a compact binary system at the second-half post-Newtonian order. Applying such equations of motion to quasi-circular orbits, we obtain the time derivative of the orbital frequency, which is consistent with the angular momentum balance equation. Numerical estimates of the phase of gravitational waves are provided for typical mass transfer rates. Our result  can be used to improve the waveforms of gravitational waves emitted by compact binaries with mass transfer. 

\end{abstract}
 
 \maketitle
 
\section{Introduction}

The mass transfer (MT) processes are widespread in compact binary systems \cite{Accretion_power,2023pbse.book.....T,Fernandez_2013,Chen_2023}. In our previous studies \cite{PhysRevD.109.123013,PhysRevD.111.043049}, we find that MT process introduces additional phase corrections to the gravitational waveforms emitted by inspiralling compact binary systems and the MT effect can be detected by space-based detectors. 
This effect becomes comparable to the effect of gravitational wave (GW) radiation at the MT rate of $10^{-6} M_\odot/\text{year}$ and in the frequency band of 10 mHz \cite{PhysRevD.111.043049}. The phase corrections can also help measure individual masses of binary white dwarfs.


Previous studies have focused on Newtonian order effects of MT in quasi-circular orbits \cite{PhysRevD.111.043049,10.1093/mnras/stab626,10.1093/mnras/stad2358}. The rate of MT typically becomes more pronounced as the binary stars approach each other \cite{10.1093/mnras/stab626,PhysRevD.109.123013}. This necessitates an investigation into the effects of MT within post-Newtonian (PN) approximations. The classical method to analytically calculating the orbital dynamics and GW signals of binary systems involves solving Einstein’s field equations by using high-order PN approximations within the low-velocity limit \cite{blanchet2024postnewtoniantheorygravitationalwaves,Blanchet2014, Gravity_PoissonWill, Maggiore:2007ulw}. In the absence of the MT effect, the post-Newtonian approximation has been extended to the 4PN order in the “even” sector and the 5.5PN order in the “odd” sector \cite{PhysRevD.37.1410, LucBlanchet_1998, Blanchet:2023sbv}.

The main purpose of this study is to derive the equations of motion of inspiralling compact binaries with MT at the 2.5PN order. The variation in mass induces additional general relativistic effects on spacetime metric and modifies the equations of motion of the binary. For the conservative parts of the binary system up to the 2PN order, we derive the equations of motion by the variation of the Fokker action \cite{blanchet2024postnewtoniantheorygravitationalwaves, PhysRevD.93.084037} with MT, which is obtained by eliminating the field degrees of freedom in the total action. We incorporate the acceleration due to the radiation reaction of GW with MT at the 2.5PN order. The variation of GW frequency is obtained in the case of quasi-circular orbits and we analyze the effect of MT at each PN order. We find that the MT effect at the 1PN order is non-negligible. Specifically, the MT effect contributes $2.68$ GW cycles at the MT rate of $10^{-5} M_{\odot}/\text{year}$ and the frequency of $0.1$ Hz at the 1PN order.

This paper is organized as follows. In Section \ref{Sec2}, we present the post-Newtonian retarded potential $V$ with MT in the metric at the 2PN order. In Section \ref{Sec3}, we derive the Fokker action with MT at the 2PN order. In Section \ref{Sec4}, we obtain the relative acceleration of compact binaries at the 2.5PN order. In Section \ref{Sec5}, we derive the variation of the frequency of GWs in the case of quasi-circular orbits and analyze the contribution of the MT effect at different PN orders. Throughout, we use the Minkowski metric $\eta_{\mu\nu}$ with signature $(-,+,+,+)$ and choose the harmonic gauge $\partial_\mu(g^{\mu\nu}\sqrt{-g})=0$.

\section{2PN Metric and potential}\label{Sec2}

We will derive the specific expressions of the metric $g_{\mu\nu}$, the energy-momentum tensor $T^{\mu\nu}$, and the PN potential $V_\mu=(V,V_i)$ with the MT effect at the 2PN order. Considering the MT, the 2PN metric takes the form as \cite{blanchet2024postnewtoniantheorygravitationalwaves}
\begin{align}
   g_{00}&=-1+\frac{2}{c^2}V-\frac{2}{c^4}V^2+\frac{8}{c^6}\left(X+V_iV^i+\frac{1}{6}V^3\right)+\mathcal{O}(c^{-8}),\\
   g_{0i}&=-\frac{4}{c^{3}}V_i-\frac{8}{c^5}R_i+\mathcal{O}(c^{-7}),\\
   g_{ij}&=\delta_{ij}\left(1+\frac{2}{c^2}V+\frac{2}{c^4}V^2\right)+\frac{4}{c^4}W_{ij}+\mathcal{O}(c^{-6}).
\end{align}
which is also applicable to the binary system with MT \cite{Gravity_PoissonWill,Blanchet2014,PhysRevD.63.062005}. This PN metric is expanded in powers of $1/c$, with $c$ the speed of light. The PN potentials $V,W,R,X$ are determined by \cite{PhysRevD.63.062005,PhysRevD.81.064004}
\begin{gather}
    \Box {V}=-4\pi G\sigma,\qquad \Box{V_i}=-4\pi G\sigma_i,\\
    \Box W_{ij}=-4\pi G(\sigma_{ij}-\delta_{ij}\sigma_{k}^{\;\;k})-\partial_i V\partial_j V,\\
    \Box R_i=-4\pi G(V\sigma_i-V_i\sigma)-2\partial^kV\partial_iV_k-\frac{3}{2}\partial_t V\partial_i V,\\
    \Box X=-4\pi G V\sigma_{i}^{\;\;i}+W_{ij}\partial^{i}\partial^j V+2V_i\partial_t\partial^i V+V\partial_t^2V+\frac{3}{2}(\partial_t V)^2-2\partial^iV_j\partial^jV_i.
\end{gather}
where $G$ is the gravitation constant. The PN potential $V$ is determined by the Energy-momentum tensor $T^{\mu\nu}$ of matter and $\sigma,\sigma_i,\sigma_{ij}$ denote the compact-support parts of the source, defined from \cite{blanchet2024postnewtoniantheorygravitationalwaves}
\begin{equation}
    \sigma=\frac{1}{c^2}(T^{00}+T^{ii}),\quad \sigma^i=\frac{1}{c}T^{0i},\quad \sigma^{ij}=T^{ij}.
\end{equation}
The indexes of $\sigma^i$ and $\sigma^{ij}$ can be lowered by the Minkowski metric $\eta_{\mu\nu}$ to get $\sigma_i$ and $\sigma_{ij}$. We need to calculate this potential from the Newtonian potential $\bigtriangleup U\equiv\partial_i\partial^i U= -4\pi G\sigma$, $\bigtriangleup U_i= -4\pi G\sigma_i$, and $\bigtriangleup U_{ij}=8\pi G\sigma_{ij}-\partial_i U\partial_j U$. The potential $U$ is related to $V$ by $\Box V=(\bigtriangleup-c^{-2}\partial_t^2)V=\bigtriangleup U $, and $\bigtriangleup\equiv\nabla^2\equiv\partial_i\partial^i$. Based on this transformation, the retarded integration $\Box^{-1}$ can be transformed to the Poisson integration $\bigtriangleup^{-1}$ as \cite{blanchet2024postnewtoniantheorygravitationalwaves} 
\begin{align}\label{VU}
    V&=U+\frac{1}{2c^2}\partial_t^2(2\bigtriangleup^{-1}U)+\frac{1}{24c^4}\partial_t^4(24\bigtriangleup^{-2}U)+\mathcal{O}(c^{-6}),\\
    V_i&=U_i+\frac{1}{2c^2}\partial_t^2\bigtriangleup^{-1}(2U_i)+\mathcal{O}(c^{-4}),
\end{align}
and the Poisson integration can be written as \cite{661cfda7-9a69-35d7-8947-df9a1ce31bd5} 
\begin{equation}\label{PoisonI}
    \bigtriangleup^{-n-1}[...]=-\frac{1}{4\pi}\int\mathrm{d}\bm{x}'\frac{1}{(2n)!}|\bm{x}-\bm{x}'|^{2n-1}[...],
\end{equation}
and $U$ can be expressed by the energy-momentum tensor as \cite{blanchet2024postnewtoniantheorygravitationalwaves}
\begin{align}\label{UT}
U(\bm{x},t)&=G\int\frac{\mathrm{d}^3 \bm{x}'}{|\bm{x}-\bm{x}'|}\left[T^{00}(\bm{x}',t)+T^{ii}(\bm{x}',t)\right],\\
    U^i(\bm{x},t)&=G\int\frac{\mathrm{d}^3 \bm{x}'}{|\bm{x}-\bm{x}'|}\left[T^{0i}(\bm{x}',t)\right],\\
    U^{ii}(\bm{x},t)&=-2G\int\frac{\mathrm{d}^3 \bm{x}'}{|\bm{x}-\bm{x}'|}\left[T^{ii}(\bm{x}',t)\right]+\frac{G^2}{c^4}\int\frac{\mathrm{d}^3 \bm{x}'}{|\bm{x}-\bm{x}'|}\left[T^{00}+T^{ii}\right](\bm{x}',t)\int\frac{\mathrm{d}^3 \bm{x}''}{|\bm{x}'-\bm{x}''|}\left[T^{00}+T^{ii}\right](\bm{x}'',t)-\frac{1}{2}U^2(\bm{x},t).
\end{align}
where $\bm{x}$ and $t$ is the position and time of the instantaneous Newtonian potential. 
We can write the energy-momentum tensor of particles as
\begin{equation}\label{Tmunu}
    T^{\mu\nu}=\frac{1}{\sqrt{-g}}\sum_{A}\gamma_A m_A v_A^\mu v_A^\nu\delta^3(\bm{x}-\bm{x}_A),
\end{equation}
where $\delta^3(\bm{x}-\bm{x}_A)$ is the three-dimension Dirac delta function, $g\equiv \text{det}(g_{\mu\nu})$, and  $\gamma$ at the 2PN order is \cite{blanchet2024postnewtoniantheorygravitationalwaves}
\begin{align}
    \gamma_A&=1+\frac{1}{c^2}\left(V_A+\frac{1}{2}v_A^2\right)+\frac{1}{c^4}\left(-4v_A^iV_{Ai}+\frac{5}{2}V_A v_A^2+\frac{1}{2}V_A^2+\frac{3}{8}v^4_A\right)+\mathcal{O}(c^{-8}), \\   
    \frac{1}{\sqrt{-g}}&=1-\frac{2}{c^2}V+\frac{2}{c^4}\left(V^2-W^{\;\;i}_{i}\right)+\frac{1}{c^6}\left(\frac{2}{3}V^3+8VW^{\;\;i}_{i}-4V^iV_i-X\right)+\mathcal{O}(c^{-8}).
\end{align}

Here, we consider masses of particles will vary with time, denoted as $m_A\equiv m_A(t)$. This implies that the variation of the Newtonian potential is not only due to the motion of the particles but also influenced by the time-dependent masses. When the total mass of the system, $M=\sum_A m_A$ is conserved, the total angular momentum is also conserved. Since the binary system represents only a local change in matter distribution, the conservation of the energy-momentum tensor ensures that the total mass remains constant \cite{Mathisson2010}. Therefore, we note that terms such as $\partial_t^2 \bigtriangleup^{-1}U$ and $\partial_t^4 \bigtriangleup^{-2}U$ in the PN potential $V$ need to be derived for the case with variation in masses. The expressions for $U,U_i,V$, and $V_i$ are derived in Appendix \ref{A}.

The extra terms caused by MT will also appear in all of the high PN order potentials in the form of the PN potential $V$. The energy-momentum
tensor $T^{\mu\nu}$ be expanded as \cite{blanchet2024postnewtoniantheorygravitationalwaves}
\begin{align}\label{T}
    T^{00}=&\sum_A m_Ac^2\delta^3(\bm{x}-\bm{x}_A)\left\{1+\frac{1}{c^2}\left(\frac{1}{2}v_A^2-U_A\right)\right.\nonumber\\
    &\left.+\frac{1}{c^4}\left(\frac{3}{8}v^4_A+\frac{13}{2}U_A^2+2(U_A)^{\;\;i}_{i}-4(U_A)_iv_A^i+\frac{3}{2}U_Av_A^2-\frac{1}{2}\partial_t^2(2\Delta^{-1}U_A)\right)\right\}+\mathcal{O}(c^{-4}),\\
T^{0i}=&\sum_A m_Ac\delta^3(\bm{x}-\bm{x}_A)\left\{v_A^i+\frac{1}{c^2}\left(\frac{1}{2}v_A^2-U_A\right)v_A^i\right\}+\mathcal{O}(c^{-3}),\\
T^{ij}=&\sum_A m_A\delta^3(\bm{x}-\bm{x}_A)\left\{v_A^iv_A^j+\frac{1}{c^2}\left(\frac{1}{2}v_A^2-U_A\right)v_A^iv_A^j\right\}+\mathcal{O}(c^{-4}).
\end{align}
Here, we have used $U_A$ instead of $V_A$, to denote the Newtonian potential at point A without the self field of particle A. $v_A$ and $r_A$ refer to the velocity and position of the particle $A$, respectively.

\section{Action and Lagrangian}\label{Sec3}
Just like the derivation of EIH equation \cite{EIH}, this section we will derive the Fokker action \cite{blanchet2024postnewtoniantheorygravitationalwaves,Damour1985, PhysRevD.93.084037} and Lagrangian with MT. Firstly, we write the total action of the particle system as
\begin{equation}\label{Action}
    S=S_g+S_m=\frac{c^3}{16\pi G}\int\mathrm{d}^4x\sqrt{-g}\left(R-\frac{1}{2}g_{\mu\nu}\Gamma^\mu\Gamma^\nu\right)+\int \sum_A m_Ac\delta^3(\bm{x}-\bm{x_A})\sqrt{-(g_{\mu\nu})_Av_A^\mu v_A^\nu}\mathrm{d}^4x,
\end{equation}
where $S_g$ is the gravitational action written in the Landau and Lifshitz  form \cite{landau2013classical}, $S_m$ is the matter action takes from the N-body system, and $R$ is the Ricci Scalar.   $\Gamma^\alpha\equiv g^{\mu\nu}\Gamma^{\alpha}_{\mu\nu}$ and $-\frac{1}{2}\sqrt{-g}g_{\mu\nu}\Gamma^\mu\Gamma^\nu$ is the standard gauge-fixing term appropriate to harmonic coordinates. $v_A^\mu\equiv \mathrm{d}x_A^\mu/\mathrm{d}t=(c,\bm{v}_A)$ is the coordinate velocity of particle $A$ and $(g_{\mu\nu})_A$ is the metric evaluated at the location of the particle $A$. $S_g$ can be written as
\begin{align}
    S_g
    &=-\frac{c^3}{16\pi G}\int\mathrm{d}^4x\sqrt{-g}\left(J+\frac{1}{2}g_{\mu\nu}\Gamma^\mu\Gamma^\nu\right),
\end{align}
where  $J=g^{\mu\nu}(\Gamma^\alpha_{\beta\nu}\Gamma^\beta_{\alpha\mu}-\Gamma^\alpha_{\mu\nu}\Gamma^\beta_{\alpha\beta})$, and $\Gamma^\alpha_{\mu\nu}=\frac{1}{2}g^{\alpha\beta}(\partial_\mu g_{\beta\nu}+\partial_\nu g_{\beta\mu}-\partial_\beta g_{\mu\nu})$ is the Christoffel symbol with metric $g_{\mu\nu}$. Then we can get $c^4\sqrt{-g}J$ and the term of gauge-fixing at the 2PN order
\begin{equation}
\begin{split}
    c^4\sqrt{-g}J
    =&-2V\nabla^2V+\frac{1}{c^2}\left\{2V\ddot{V}+8V^i\nabla^2V_i\right\}\\
    &+\frac{1}{c^4}\left\{-12V^2\ddot{V}+8W^{i}_{i}\ddot{V}+32R_i\nabla^2 V^i+16W^{i}_{i}\ddot{V}-16\dot{V}^i\partial^kW_{ik}+16VV_i\nabla^2V^i+4W^{i}_{i}V_{,k}V^{,k}\right.\\
    &\qquad\quad-8W^{j}_{i}V^{,i}V_{,j}+32R_i\nabla^2V^i+32R^{k}_{,k}V^{i}_{,i}-32V_iV_jV^{,i,j}-32V_jV^i_{,i}V_{,j}-16X\nabla^2V\\
    &\qquad\quad\left.-8W^{j}_{i}\nabla^2W^{j}_{i}+8W^{i}_{i}\nabla^2W^{j}_{j}-8W^{k}_{k}W_{ij}^{,i,j}-8W_{j}^{i,j}W_{ik}^{,k}\right\}+\mathcal{O}(c^{-6})\\
    &+\text{the terms of total derivative}\left[\partial_t(...),\partial_i(...)\right],
    \end{split}
\end{equation}

\begin{equation}
\begin{split}
    c^4\frac{1}{2}\sqrt{-g}g_{\mu\nu}\Gamma^\mu\Gamma^\nu
    =&\frac{1}{c^4}\left\{2V_i\nabla^2V^i+2V_{i,j}V^{i,j}+16\dot{V}^i W_{i,j}^j-8\dot{V}^i W_{j,i}^j-8W_{i,j}^jW_{,k}^{ik}+8W_{i,j}^jW_{k,i}^k-2W_{j,i}^jW_{k,i}^k\right\}\\
    &+\mathcal{O}(c^{-6})+\text{the terms of total derivative}\left[\partial_t(...),\partial_i(...)\right].
    \end{split}
\end{equation}
Here $(...)_{,\mu}\equiv \partial_\mu(...)$ and $\partial_0=c^{-1}\partial_t$. We have used the relation $V^i_{,i}+cV_{,0}=0$ which is due to the continuity equation.  The contribution of terms of gauge-fixing starts from the 2PN order. From Eq.\eqref{Vtm} we find that the effect caused by the variation in mass comes from the order of $c^{-2}$ in the PN potential $V$. Terms with the mass variation at the 2PN order can be calculated by the 1PN potential. Other 2PN potentials can be calculated as the classical PN method since their effects incorporate the variation in mass starting from the 3PN order.


Substituting the energy-momentum tensor and the PN potentials, we obtain the 2PN  action with variation in mass:
\begin{align}
    S=&\int\mathrm{d}t\sum_A m_A\bigg\{\frac{v_A^2}{2}+\frac{V_A}{2}+\frac{1}{c^2}\left(\frac{3}{4}V_Av_A^2+\frac{v_A^4}{8}-2v_A^iV_{Ai}\right)\notag\\
    &\qquad\qquad\qquad-\sum_{B\neq A}\frac{G }{c^4}\left(\frac{1}{4}\ddot{m}_B r_{AB}+\frac{1}{2}\dot{m}_B(\bm{v}_B\cdot\bm{n}_{BA})\right)V_A\bigg\}+S_{2PN0}+\mathcal{O}(c^{-6}).
\end{align}
Here, MT terms appear at the 2PN order. There are implicit corrections due to MT in the PN potential $V$ which we will expand in the Lagrangian $\mathcal{L}$. $S_{2PN0}$ denotes the 2PN terms in the action  without the MT effect. Variation of $S$ will yield the equations of motion. We derive the N-body system Lagrangian with MT in Appendix \ref{AppB}.

The Lagrangian in the case of a binary system, including only two different particles A and B is
\begin{equation}\label{Lagrangian}
    \mathcal{L}=\mathcal{L}_0+\mathcal{L}_{MT}+\mathcal{O}(c^{-6})+\mathcal{O}(\dot{m}^{-2}),
\end{equation}
where
\begin{align}
    \mathcal{L}_0&=\frac{1}{2}m_Av_A^2+\frac{1}{2}m_Bv_B^2+\frac{Gm_Am_B}{r_{AB}}\notag\\
    &+\frac{1}{c^2}\Bigg\{m_A\frac{v_A^4}{8}+m_B\frac{v_B^4}{8}+\frac{Gm_Am_B}{r_{AB}}\Bigg[\frac{3}{2}v_A^2+\frac{3}{2}v_B^2-\frac{7}{2}\bm{v}_A\cdot\bm{v}_B-\frac{1}{2}(\bm{v}_A\cdot\bm{n}_{AB})(\bm{v}_B\cdot\bm{n}_{AB})-\frac{1}{2}\frac{GM}{r_{AB}}\Bigg]\Bigg\}\notag\\
    &+\frac{1}{c^4}\Bigg\{m_A\frac{v_A^6}{16}+m_B\frac{v_B^6}{16}+\frac{G^3M^2m_Am_B}{2r_{AB}^3}+\frac{15G^3m_A^2m_B^2}{4r_{AB}^3}+\frac{Gm_Am_B}{r_{AB}}\Bigg[\frac{3}{8}(\bm{v}_A\cdot\bm{n}_{AB})^2(\bm{v}_B\cdot\bm{n}_{AB})^2+\frac{7}{8}v_A^4+\frac{7}{8}v_B^4\notag\\
    &\quad-\frac{7}{8}(\bm{v}_B\cdot\bm{n}_{AB})^2v_A^2-\frac{7}{8}(\bm{v}_A\cdot\bm{n}_{BA})^2v_B^2-2(v_A^2+v_B^2)(\bm{v}_A\cdot\bm{v}_{B})+\frac{3}{2}(\bm{v}_A\cdot\bm{n}_{AB})(\bm{v}_B\cdot\bm{n}_{AB})(\bm{v}_A\cdot\bm{v}_{B})\notag\\
    &\quad+\frac{1}{4}(\bm{v}_A\cdot\bm{v}_{B})^2+\frac{15}{8}v_A^2v_B^2+r_{AB}\bigg[-\frac{7}{4}(\bm{a}_A\cdot\bm{v}_{B})(\bm{v}_B\cdot\bm{n}_{AB})-\frac{1}{8}(\bm{a}_A\cdot\bm{n}_{AB})(\bm{v}_B\cdot\bm{n}_{AB})^2+\frac{7}{8}(\bm{a}_A\cdot\bm{n}_{AB})v_B^2\notag\\
    &\quad\qquad\qquad\qquad\qquad\qquad\qquad\qquad\;\;\;-\frac{7}{4}(\bm{a}_B\cdot\bm{v}_{A})(\bm{v}_A\cdot\bm{n}_{BA})-\frac{1}{8}(\bm{a}_B\cdot\bm{n}_{BA})(\bm{v}_A\cdot\bm{n}_{BA})^2+\frac{7}{8}(\bm{a}_B\cdot\bm{n}_{BA})v_A^2\bigg]\notag\\
    &\quad+\frac{Gm_A}{r_{AB}}\bigg(\frac{7}{2}(\bm{v}_A\cdot\bm{n}_{AB})^2-\frac{7}{2}(\bm{v}_B\cdot\bm{n}_{AB})(\bm{v}_A\cdot\bm{n}_{AB})+\frac{1}{2}(\bm{v}_B\cdot\bm{n}_{AB})^2+\frac{1}{4}v_A^2-\frac{7}{4}(\bm{v}_A\cdot\bm{v}_{B})+\frac{7}{4}v_B^2\bigg)\notag\\
    &\quad+\frac{Gm_B}{r_{AB}}\bigg(\frac{7}{2}(\bm{v}_B\cdot\bm{n}_{BA})^2-\frac{7}{2}(\bm{v}_A\cdot\bm{n}_{BA})(\bm{v}_B\cdot\bm{n}_{BA})+\frac{1}{2}(\bm{v}_A\cdot\bm{n}_{BA})^2+\frac{1}{4}v_B^2-\frac{7}{4}(\bm{v}_B\cdot\bm{v}_{A})+\frac{7}{4}v_A^2\bigg)\Bigg]\Bigg\},
    \end{align}
    \begin{align}
    \mathcal{L}_{MT}&=\frac{1}{c^2}\frac{Gm_Am_B}{r_{AB}}\Bigg\{\frac{1}{4}\left(\frac{\dot{m}_B}{m_B}(\bm{v}_B\cdot\bm{n}_{BA})+\frac{\dot{m}_A}{m_A}(\bm{v}_A\cdot\bm{n}_{AB})\right)r_{AB}\Bigg\}\notag\\
   &+\frac{1}{c^4}\frac{Gm_Am_B}{r_{AB}}\Bigg\{\frac{\dot{m}_A}{m_A} \bigg[r_{AB}^2\bigg(2(\bm{v}_B\cdot\bm{a}_A)+\frac{5}{4} (\bm{a}_A\cdot\bm{v}_A)+\frac{1}{2} (\bm{a}_A\cdot\bm{n}_{AB}) (\bm{v}_A\cdot\bm{n}_{AB})\bigg)\notag\\
      &\qquad\qquad+r_{AB}\bigg(\frac{3}{8}(\bm{v}_A\cdot\bm{n}_{AB})v_B^2+\frac{3}{2} (\bm{v}_A\cdot\bm{n}_{AB}) v_A^2+2(\bm{v}_{A}\cdot\bm{n}_{AB})(\bm{v}_A\cdot\bm{v}_B)-\frac{1}{4}  (\bm{v}_A\cdot\bm{n}_{AB})^3\bigg)\bigg]\notag\\
            &\quad+\frac{\dot{m}_B}{m_B} \bigg[r_{AB}^2\bigg(2(\bm{v}_A\cdot\bm{a}_B)+\frac{5}{4} (\bm{a}_B\cdot\bm{v}_B)+\frac{1}{2} (\bm{a}_B\cdot\bm{n}_{BA}) (\bm{v}_B\cdot\bm{n}_{BA})\bigg)\notag\\
      &\qquad\qquad+r_{AB}\bigg(\frac{3}{8}(\bm{v}_B\cdot\bm{n}_{BA})v_A^2+\frac{3}{2} (\bm{v}_B\cdot\bm{n}_{BA}) v_B^2+2(\bm{v}_{B}\cdot\bm{n}_{BA})(\bm{v}_A\cdot\bm{v}_B)-\frac{1}{4}  (\bm{v}_B\cdot\bm{n}_{BA})^3\bigg)\bigg]\notag\\
    &\quad+\frac{G m_A}{r_{AB}}\Bigg[\frac{\dot{m}_A}{m_A} r_{AB}\bigg( -(\bm{v}_B\cdot\bm{n}_{BA})+\frac{3}{4}(\bm{v}_A\cdot\bm{n}_{BA})\bigg)+\frac{\dot{m}_B}{m_B} \left(\frac{3}{8} r_{AB}(\bm{v}_{B}\cdot\bm{n}_{BA})+\frac{1}{8} r_{AB}(\bm{v}_{A}\cdot\bm{n}_{BA})\right)\Bigg]\notag\\
        &\quad+\frac{G m_B}{r_{AB}}\Bigg[\frac{\dot{m}_B}{m_B}r_{AB} \bigg(- (\bm{v}_A\cdot\bm{n}_{AB})+\frac{3}{4}(\bm{v}_B\cdot\bm{n}_{AB})\bigg) +\frac{\dot{m}_A}{m_A} \left(\frac{1}{8} r_{AB}(\bm{v}_{B}\cdot\bm{n}_{AB})+\frac{3}{8} r_{AB}(\bm{v}_{A}\cdot\bm{n}_{AB})\right)\Bigg]\Bigg\}.
\end{align}
Through our derivation, we found that the Lagrangian $\mathcal{L}_0$ without MT is consistent with the Lagrangian in Eq.(4.1) in \cite{VanessaCdeAndrade_2001}. Meanwhile, additional effects caused by MT in Lagrangian $\mathcal{L}_{MT}$ are obtained. In non-extreme cases (excluding violent accretion or extreme eccentricity), the MT rate is typically considered to be small. At a frequency of 10 mHz and a MT rate of $10^{-5}M_\odot/\text{year}$, the  contribution of $\dot{m}$ at the Newtonian-order is comparable to that of GW radiation at the 2.5 PN order \cite{PhysRevD.111.043049} and the contribution of $\dot{m}^2$ is much smaller than that of GW at the 2.5 PN order. Therefore, terms of the order $\dot{m}^2$ can be regarded as a higher-order  quantity and be neglected in the calculation. Additionally, in cases without extreme eccentricity, the variation in the MT rate is not significant, implying that the second derivative of masses $\ddot{m}$ can also be treated as a small quantity and neglected.

The Lagrangian in the center-of mass frame is obtained by the transformation in Appendix \ref{AppD}. Terms without MT are the same as the guess-work in Eq.(4.2) in \cite{LucBlanchet_2003} and the result is  
\begin{align}\label{CoML}
    \mathcal{L}&=\frac{1}{2}M^2\eta v^2+\frac{G M^2\eta}{r}\notag\\
    &+\frac{1}{c^2}\bigg\{-\frac{G^2 M^3 \eta }{2 r^2}+\frac{G M^2\eta }{r} \left(\frac{1}{2} \eta v^2+\frac{3}{2} v^2+\frac{1}{2}\eta \dot{r}^2\right)-\frac{3}{8} M \eta ^2 v^4+\frac{1}{8} M \eta  v^4-\frac{1}{4} G M \dot{r} \Delta \chi \bigg\}\notag\\
    &+\frac{1}{c^4}\bigg\{M\eta v^6\bigg(\frac{13}{16}  \eta ^2-\frac{7}{16} \eta+\frac{1}{16}\bigg)+\frac{G^3 M^4\eta}{r^3} \left(\frac{15  }{4}\eta+\frac{1}{2}\right)\notag\\
    &\quad+\frac{GM^2\eta}{r} \bigg(\frac{7}{8}v^4-\frac{5 }{4}\eta  v^4-\frac{9 }{8}\eta ^2 v^4+\frac{1}{4 }\eta  \dot{r}^2 v^2-\frac{5}{4 }\eta ^2 \dot{r}^2 v^2+\frac{3 }{8}\eta ^2 \dot{r}^4+\frac{7}{8}r (\bm{a}\cdot\bm{n}) \eta  v^2-\frac{1}{8} r (\bm{a}\cdot\bm{n}) \eta  \dot{r}^2-\frac{7}{4} r (\bm{a}\cdot\bm{v}) \eta  \dot{r}\bigg)\notag\\
    &\quad+\frac{G^2M^3\eta}{r^2} \bigg(\frac{v^2 \eta ^2}{2 }+\frac{3 \eta ^2 \dot{r}^2 }{2 }+\frac{41 \eta  \dot{r}^2}{8 }+\frac{ \dot{r}^2}{2}+\frac{7 v^2}{4 }-\frac{27 v^2 \eta }{8 }\bigg)\notag\\
    &\quad+\Delta \chi \bigg[GM \left(\bigg(\frac{13}{4} \eta-\frac{5}{4}\bigg) r (\bm{a}\cdot\bm{v})+\bigg(\frac{1}{2}  \eta -\frac{1}{2} \bigg) r \dot{r} (\bm{a}\cdot\bm{n})+\bigg(-\frac{3}{8}  \eta+\frac{1}{4}\bigg)  \dot{r}^3 +\bigg(\frac{41}{8}  \eta -\frac{3}{2}\bigg)  v^2 \dot{r}\right)-\frac{1}{2}\frac{G^2M^2}{r}\dot{r} \bigg]\bigg\}\notag\\
    &\qquad\qquad+\mathcal{O}(c^{-6}).
\end{align}
This form is much more simpler than Eq.\eqref{Lagrangian}. We will use this Lagrangian to derive angular momentum and angular acceleration in the next section. One peculiarity is that all MT terms include the dimensionless relative mass difference\footnote{Note that the symbols for the relative mass difference $\Delta$ and the Laplace operator $\bigtriangleup$ are different.} $\Delta$  and the MT rate $\chi$.

\section{Equations of motion}\label{Sec4}
\subsection{The 2PN acceleration in harmonic-coordinate}
Because of the appearance of the acceleration in the 2PN Lagrangian in Eq.\eqref{Lagrangian}, we need to calculate the derivative of $\mathcal{L}$ with acceleration $a_A^i$, the Euler-Lagrange equation is
\begin{equation}\label{E-L}
    \frac{\partial \mathcal{L}}{\partial x_{A}^i}-\frac{\mathrm{d}}{\mathrm{d}t}\left(\frac{\partial\mathcal{L}}{\partial v_A^i}\right)+\frac{\mathrm{d}^2}{\mathrm{d}t^2}\left(\frac{\partial\mathcal{L}}{\partial a_A^i}\right)=-\mathcal{F},
\end{equation}
where $\mathcal{F}$ is the non-conservative terms, such as the 2.5PN order GW radiation and MT dissipation. We will incorporate the GW radiation in the next subsection. In some symmetry situations there would be an accretion disk around the primary star while the orbit angular momentum will translate to the spin angular momentum of accretion disk \cite{PhysRevD.111.043049,1988ApJ...332..193V}. In following calculations we just take $\mathcal{F}=0$ to derive the conservative terms of orbit at the 1PN and 2PN orders. Substituting $\mathcal{L}$ into the Euler-Lagrange equation, we get the acceleration $a_A$ as
\begin{equation}\label{aA}
    \bm{a}_A=-\frac{Gm_B}{r^2}\Big[\left(1+\mathcal{A}_0+\mathcal{A}_{MT}\right)\bm{n}+\left(\mathcal{B}_0+\mathcal{B}_{MT}\right)\bm{v}\Big]+\Big(-\frac{\dot{m}_A}{m_A}+\mathcal{C}_{MT}\Big)\bm{v}_A+\mathcal{O}(c^{-6}),
\end{equation}
where $\bm{n}=\bm{x}/r$, orbital separation and relative position $r\equiv r_{AB}=|\bm{x}|=|\bm{x}_A-\bm{x}_B|$, $\dot{r}=\bm{v}\cdot\bm{n}$ and relative velocity $\bm{v}=\bm{v}_A-\bm{v}_B=\mathrm{d}\bm{x}/\mathrm{d}t$.  The coefficients are given by
\begin{align}
    \mathcal{A}_0&=\frac{1}{c^2}\bigg\{-5\frac{Gm_A}{r}-4\frac{Gm_B}{r}-\frac{3}{2}(\bm{v}_B\cdot\bm{n})^2-4\bm{v}_A\cdot\bm{v}_B+v_A^2+2v_B^2\bigg\}\notag\\
    &+\frac{1}{c^4}\bigg\{\frac{57}{4}\frac{G^2m_A^2}{r^2}+\frac{69}{2}\frac{G^2m_Am_B}{r^2}+9\frac{G^2m_B^2}{r^2}+\frac{15}{8}(\bm{v}_B\cdot\bm{n})^4-\frac{3}{2}(\bm{v}_B\cdot\bm{n})^2v_A^2+6(\bm{v}_B\cdot\bm{n})^2(\bm{v}_A\cdot\bm{v}_B)\notag\\
    &\qquad-\frac{9}{2}(\bm{v}_B\cdot\bm{n})^2v_B^2+2(\bm{v}_A\cdot\bm{v}_B)^2-2v_B^4-4(\bm{v}_A\cdot\bm{v}_B)v_B^2\notag\\
    &\qquad-\frac{Gm_B}{r}\bigg[2(\bm{v}_A\cdot\bm{n})^2-4(\bm{v}_A\cdot\bm{n})(\bm{v}_B\cdot\bm{n})-6(\bm{v}_B\cdot\bm{n})^2-8(\bm{v}_A\cdot\bm{v}_B)+4v_B^2\bigg]\notag\\
    &\qquad-\frac{Gm_A}{r}\bigg[\frac{39}{2}(\bm{v}_A\cdot\bm{n})^2-39(\bm{v}_A\cdot\bm{n})(\bm{v}_B\cdot\bm{n})+\frac{17}{2}(\bm{v}_B\cdot\bm{n})^2-\frac{15}{4}v_A^2-\frac{5}{2}(\bm{v}_A\cdot\bm{v}_B)+\frac{5}{4}v_B^2\bigg]\bigg\},\\
    \mathcal{A}_{MT}&=\frac{1}{c^2}\bigg\{-\frac{3}{4} r  (\bm{v}_B\cdot\bm{n})\frac{\dot{m}_A}{m_A}+{\frac{1}{4}r (\bm{v}_B\cdot\bm{n}) \frac{\dot{m}_B}{m_B}}\bigg\}\notag\\
    &+\frac{1}{c^4}\bigg\{r\frac{\dot{m}_A}{m_A} \bigg[\frac{G m_A}{r} \Big(-\frac{11}{2} (\bm{v}_B\cdot\bm{n})+\frac{145}{8} (\bm{v}_A\cdot\bm{n})\Big)+\frac{G m_B}{r} \Big(-\frac{3}{2} (\bm{v}_A\cdot\bm{n}) -\frac{3}{2} (\bm{v}_B\cdot\bm{n})\Big)\notag\\
    &\qquad\qquad+\frac{7}{8} (\bm{v}_B\cdot\bm{n})v_A^2+\frac{1}{2} (\bm{v}_A\cdot\bm{n}) v_B^2-\frac{15}{8} (\bm{v}_B\cdot\bm{n}) v_B^2-\frac{3}{2} (\bm{v}_A\cdot\bm{n}) (\bm{v}_B\cdot\bm{n})^2+\frac{3}{4} (\bm{v}_A\cdot\bm{n})^2 (\bm{v}_B\cdot\bm{n})+\frac{3}{4} (\bm{v}_B\cdot\bm{n})^3\notag\\
    &\qquad\qquad+5 (\bm{v}_A\cdot\bm{v}_B) (\bm{v}_B\cdot\bm{n})-\frac{1}{2} (\bm{v}_A\cdot\bm{n}) (\bm{v}_A\cdot\bm{v}_B)\bigg]\notag\\
    &\qquad+r\frac{\dot{m}_B}{m_B} \bigg[\frac{Gm_A}{r} \Big(\frac{13}{4}(\bm{v}_B\cdot\bm{n})-\frac{29}{8} (\bm{v}_A\cdot\bm{n})\Big)+\frac{Gm_B}{r} \Big(2 (\bm{v}_A\cdot\bm{n})-\frac{15}{4} (\bm{v}_B\cdot\bm{n})\Big)\notag\\
    &\qquad\qquad+\frac{1}{8} v_A^2 (\bm{v}_B\cdot\bm{n})+\frac{1}{4} (\bm{v}_A\cdot\bm{n}) v_B^2+\frac{11}{8} (\bm{v}_B\cdot\bm{n}) v_B^2-\frac{3}{4} (\bm{v}_A\cdot\bm{n}) (\bm{v}_B\cdot\bm{n})^2+\frac{3}{8} (\bm{v}_A\cdot\bm{n})^2 (\bm{v}_B\cdot\bm{n})-\frac{3}{8} (\bm{v}_B\cdot\bm{n})^3\notag\\
    &\qquad\qquad-\frac{1}{4}  (\bm{v}_A\cdot\bm{n}) (\bm{v}_A\cdot\bm{v}_B)-\frac{11}{2}  (\bm{v}_A\cdot\bm{v}_B) (\bm{v}_B\cdot\bm{n})\bigg]\bigg\},\\
    \mathcal{B}_0&=\frac{1}{c^2}\bigg\{-4(\bm{v}_A\cdot\bm{n})+3(\bm{v}_B\cdot\bm{n})\bigg\}\notag\\
    &+\frac{1}{c^4}\bigg\{\frac{Gm_B}{r}\Big[2(\bm{v}_A\cdot\bm{n})+2(\bm{v}_B\cdot\bm{n})\Big]+\frac{Gm_A}{r}\Big[\frac{63}{4}(\bm{v}_A\cdot\bm{n})-\frac{55}{4}(\bm{v}_B\cdot\bm{n})\Big]-4(\bm{v}_A\cdot\bm{n})v_B^2+5(\bm{v}_B\cdot\bm{n})v_B^2\notag\\
    &\qquad+6(\bm{v}_A\cdot\bm{n})(\bm{v}_B\cdot\bm{n})^2-\frac{9}{2}(\bm{v}_B\cdot\bm{n})^3-(\bm{v}_B\cdot\bm{n})v_A^2+4(\bm{v}_A\cdot\bm{n})(\bm{v}_A\cdot\bm{v}_B)-4(\bm{v}_B\cdot\bm{n})(\bm{v}_A\cdot\bm{v}_B)\bigg\},\\
    \mathcal{B}_{MT}&=\frac{1}{c^2}\bigg\{{\frac{13}{4}r\frac{\dot{m}_A}{m_A}+\frac{1}{4}r\frac{\dot{m}_B}{m_B}}\bigg\}\notag\\
    &+\frac{1}{c^4}\bigg\{r\frac{\dot{m}_A}{m_A} \bigg[-16\frac{Gm_A}{r}-7 \frac{Gm_B}{r}+3 (\bm{v}_A\cdot\bm{n}) (\bm{v}_B\cdot\bm{n})+ (\bm{v}_A\cdot\bm{v}_B)+\frac{1}{4} (\bm{v}_A\cdot\bm{n})^2-\frac{21}{4} (\bm{v}_B\cdot\bm{n})^2-\frac{25}{8} v_A^2+\frac{49}{8} v_B^2\bigg]\notag\\
    &+r\frac{\dot{m}_B}{m_B} \bigg[7\frac{Gm_A}{r}- \frac{Gm_B}{r}-\frac{5}{2} (\bm{v}_A\cdot\bm{n}) (\bm{v}_B\cdot\bm{n})+\frac{1}{2}(\bm{v}_A\cdot\bm{v}_B)-\frac{15}{8} (\bm{v}_A\cdot\bm{n})^2+\frac{61}{8} (\bm{v}_B\cdot\bm{n})^2+\frac{17}{8} v_A^2-\frac{53}{8} v_B^2\bigg]\bigg\},\\
    \mathcal{C}_{MT}&=\frac{1}{c^2}\bigg\{{v_A^2\frac{\dot{m}_A}{m_A}+\frac{Gm_B}{r}\Big(\frac{13}{4}\frac{\dot{m}_A}{m_A}-\frac{11}{4}\frac{\dot{m}_B}{m_B}\Big)}\bigg\}\notag\\
    &+\frac{1}{c^4}\bigg\{r\frac{\dot{m}_A}{m_A} \bigg[\frac{Gm_B}{r} \Big(\frac{11}{4} (\bm{v}_A\cdot\bm{n}) (\bm{v}_B\cdot\bm{n})-\frac{27}{4} (\bm{v}_A\cdot\bm{v}_B)+(\bm{v}_A\cdot\bm{n})^2-\frac{13}{2} (\bm{v}_B\cdot\bm{n})^2+\frac{1}{8}v_A^2+\frac{59}{8} v_B^2\Big)\notag\\
    &\qquad\qquad-\frac{105 }{8}\frac{G^2m_A m_B}{r^2}-\frac{25}{4}\frac{G^2 m_B^2}{ r^2}\bigg]\notag\\
    &\qquad+r\frac{\dot{m}_B}{m_B} \bigg[\frac{Gm_B}{r} \Big( -\frac{5}{4} (\bm{v}_A\cdot\bm{n}) (\bm{v}_B\cdot\bm{n})-\frac{3}{4} (\bm{v}_A\cdot\bm{v}_B)-\frac{15}{8} (\bm{v}_A\cdot\bm{n})^2+\frac{41 }{8} (\bm{v}_B\cdot\bm{n})^2+\frac{25}{8} v_A^2-\frac{29}{8} v_B^2\Big)\notag\\
    &\qquad\qquad+\frac{133}{8} \frac{G^2 m_A m_B}{r^2}+\frac{G^2m_B^2}{r^2}\bigg]\bigg\}.
\end{align}
The subscript $ MT $ denotes terms arising from the contribution of variation in mass, and $ \bm{a}_B $ can be obtained by symmetry as $ (A \leftrightarrow B) $. In the above derivation, we do not impose specific evolution forms for the binary component masses; instead, we treat $ \ddot{m} $ as a small quantity and neglect it under non-extreme conditions. Total mass conservation requires$ \dot{m}_A = -\dot{m}_B\equiv \chi $. 

It is important to note that while the accelerations $ \bm{a}_A $ and $ \bm{a}_B $ are derived within a conservative framework, this does not imply orbits are strictly circular or elliptical. The system includes orbital precession due to general relativistic effects and orbital expansion caused by the MT effect. The contribution of the MT effect on the orbit is evident starting from the Newtonian order \cite{PhysRevD.111.043049}.


\subsection{2.5PN GW back reaction}
We now incorporate the GW back reaction into the equations of motion. In the harmonic gauge and the center-of-mass frame, the 2.5PN acceleration due to the GW back reaction is given by \cite{Gravity_PoissonWill}:
\begin{equation}\label{a25PN}
    a^i_{2.5PN}=\frac{G}{c^5}\left[-\frac{GM}{r^2}\left(3{\mathop{I}^{(3)}}_{jk}n^jn^k+\frac{1}{3}\mathop{I}^{(3)}{}^{k}_k\right)n^i+2\mathop{I}^{(4)}{}^i_kv^k+\frac{3}{5}\left(\mathop{I}^{(5)}{}^i_k-\frac{1}{3}\delta^i_k\mathop{I}^{(5)}{}^j_j\right)r^k\right],
\end{equation}
where the top scripts $(3),(4)$ are the 3rd or 4th derivative with time. The time derivative of the mass quadrupole with MT, $I^{jk}_{MT}=\mu r^jr^k$ can be written as
\begin{align}
    \dot{I}^{jk}_{MT}=&\dot{I}^{jk}_0+\dot{\mu}r^jr^k,\\
    \ddot{I}^{jk}_{MT}=&\ddot{I}^{jk}_0+2r(n^jv^k+n^kv^j)\dot{\mu}+\mathcal{O}(\dot{\mu}^2),\\
    \mathop{I}^{(3)}{}^{jk}_{MT}=& \mathop{I}^{(3)}{}^{jk}_0-4\frac{GM}{r}\dot{\mu}n^jn^k+\mathcal{O}(\dot{\mu}^2),\\
    \mathop{I}^{(4)}{}^{jk}_{MT}=& \mathop{I}^{(4)}{}^{jk}_0-\frac{GM}{r^2}\bigg\{4\bigg[(n^jv^k+n^kv^j)-3\dot{r}n^kn^j\bigg]\dot{\mu}\bigg\}+\mathcal{O}(\dot{\mu}^2),\\
    \begin{split}
         \mathop{I}^{(5)}{}^{jk}_{MT}=& \mathop{I}^{(5)}{}^{jk}_0+2\frac{GM}{r^3}\left\{\left[\frac{GM}{r}n^jn^k-3\left(v^2n^jn^k+\dot{r}(n^jv^k+n^kv^j)-5\dot{r}^2n^jn^k\right)\right]\dot{\mu}\right\}+\mathcal{O}(\dot{\mu}^2),
    \end{split}
\end{align}
where $\mu=m_Am_B/M$ is the reduced mass, $I^{jk}_0$ denotes the mass quadrupole without MT. Here considering only at the 2.5PN order, we use the Newtonian order's relation between relative distance $\bm{r}$ and the distances from stars A and B to the center of mass $\bm{r}_A,\bm{r}_B$ and between relative velocity $\bm{v}$ and the respective velocities of the stars $\bm{v}_A,\bm{v}_B$ as
\begin{equation}\label{vAB0}
    \bm{v}_A=\frac{m_B}{M}\bm{v},\qquad\bm{v}_B=-\frac{m_A}{M}\bm{v}.
\end{equation}
 The extension of Eq.\eqref{vAB0} to the 1PN and 2PN orders are derived in Appendix \ref{AppD}, we will use this relation in the next subsection. Substituting above formulas into Eq.\eqref{a25PN} and getting
\begin{equation}
\begin{split}
        \bm{a}_{2.5PN}=&\frac{1}{c^5}\frac{G^2M^2}{r^3}\Bigg\{\left[\frac{24}{5}\eta v^2\dot{r}+\frac{136}{15}\frac{GM}{r}\eta\dot{r}+\Big({-\frac{188}{15} \frac{GM}{r} +\frac{28}{5}  v^2-\frac{66}{5}  \dot{r}^2\Big)r\frac{\Delta\chi}{M}}\right]\bm{n}\\
        &\qquad\qquad+\left[-\frac{8}{5}\eta v^2-\frac{24}{5}\frac{GM}{r}\eta{+\frac{22}{5} r \frac{\Delta \chi}{M}  \dot{r}}\right]\bm{v}\Bigg\},
\end{split}
\end{equation}
where $\Delta=(m_A-m_B)/M$ is the dimensionless relative mass difference, $\eta=\mu/M$ is the symmetry mass ratio and $\chi=\dot{m_A}=-\dot{m_B}$ is the MT rate. 

\subsection{Center-of-mass frame}
We now derive the 2.5PN order equations of motion in the center-of-mass frame \cite{LucBlanchet_2003}. The specific derivation of the transformation is shown in Appendix \ref{AppD}.
Transforming the relative acceleration $\bm{a}=\bm{a}_A-\bm{a}_B$ into the center-of-mass frame, we have
\begin{equation}\label{CoMacceleration}
    \bm{a}=-\frac{GM}{r^2}\bigg[\left(1+\Tilde{\mathcal{A}}_0+\Tilde{\mathcal{A}}_{MT}\frac{\Delta }{\mu}r\chi\right)\bm{n}+\left(\Tilde{\mathcal{B}}_0+\Tilde{\mathcal{B}}_{MT}\frac{\Delta}{\mu}r\chi\right)\bm{v}\bigg]+\frac{\Delta}{\mu}\left(1+\mathcal{C}_{MT}\right)\chi\bm{v}+\mathcal{O}(c^{-6}),
\end{equation}
where
\begin{align}
    \Tilde{\mathcal{A}}_0&=\frac{1}{c^2}\bigg\{-\frac{3}{2}\dot{r}^2\eta+v^2+3\eta v^2-\frac{GM}{r}(4+2\eta)\bigg\}\notag\\
    &\,+\frac{1}{c^4}\bigg\{\frac{15}{8}\dot{r}^4\eta-\frac{45}{8}\dot{r}^4\eta^2-\frac{9}{2}\dot{r}^2\eta v^2+6\dot{r}^2\eta^2 v^2+3\eta v^4-4\eta^2 v^4+\frac{G^2M^2}{r^2}\left(9+\frac{87}{4}\eta\right)\notag\\
    &\qquad\quad+\frac{GM}{r}\left(-2\dot{r}^2-25\dot{r}^2\eta-2\dot{r}^2\eta^2-\frac{13}{2}\eta v^2+2\eta^2 v^2\right)\bigg\}\notag\\
    &\,+\frac{1}{c^5}\bigg\{-\frac{24}{5}\frac{GM}{r}\dot{r}\eta v^2-\frac{136}{15}\frac{G^2M^2}{r^2}\dot{r}\eta\bigg\},\\
    \Tilde{\mathcal{A}}_{MT}&=\frac{1}{c^2}\bigg\{- \eta \dot{r}\bigg\}+\frac{1}{c^4}\bigg\{\frac{3}{2} \eta  \dot{r}^3+\frac{G M}{r}\left(-3 \eta ^2 \dot{r}-\frac{69}{4} \eta  \dot{r}+\frac{3}{2} \dot{r}\right)\bigg\}\notag\\
    &\,+\frac{1}{c^5}\bigg\{-\frac{188}{15} \frac{G^2M^2} {r^2} \eta+\frac{GM}{r}\Big(-\frac{66}{5} \eta  \dot{r}^2+\frac{28}{5} \eta  v^2\Big)\bigg\},\\
    \Tilde{\mathcal{B}}_0&=\frac{1}{c^2}\bigg\{-4\dot{r}+2\dot{r}\eta\bigg\}+\frac{1}{c^4}\bigg\{\frac{9}{2}\dot{r}^3\eta+3\dot{r}^3\eta^2-\frac{15}{2}\dot{r}\eta v^2-2\dot{r}\eta^2v^2+\frac{GM}{r}\left(2\dot{r}+\frac{41}{2}\dot{r}\eta+4\dot{r}\eta^2\right)\bigg\}\notag\\
    &\,+\frac{1}{c^5}\bigg\{\frac{8}{5}\frac{GM}{r}\eta v^2+\frac{24}{5}\frac{G^2M^2}{r^2}\eta\bigg\},\\
    \Tilde{\mathcal{B}}_{MT}&=\frac{1}{c^2}\bigg\{-1\bigg\}+\frac{1}{c^4}\bigg\{-3 \eta ^2 \dot{r}^2-6 \eta  \dot{r}^2+\frac{3}{4} \dot{r}^2-3 \eta ^2 v^2+6 \eta  v^2+\frac{13}{4} v^2+\frac{GM}{r} \left(- \eta ^2+\frac{29}{4} \eta +\frac{3}{4} \right)\bigg\}\notag\\
    &\,+\frac{1}{c^5}\bigg\{\frac{22}{5} \eta  \dot{r}\bigg\},\\
    \Tilde{\mathcal{C}}_{MT}&=\frac{1}{c^2}\bigg\{-v^2 + \frac{3}{2} v^2 \eta\bigg\}+\frac{1}{c^4}\bigg\{\frac{9}{8} \eta  v^4-3 \eta ^2 v^4\bigg\}.
\end{align}

The relative mass difference, $\Delta$, appears alongside $\chi$ in MT terms.
In classical cases without MT, all terms related to mass in the acceleration are expressed in terms of $\mu,\eta$ and $M$, including the chirp mass $\mathcal{M}_c=(m_Am_B)^{3/5}/M^{1/5}$ in the GW strain. In the low-frequency band, additional terms involve $\Delta$ and $\chi$ destroyed the degeneracy of mass in accretion binaries by influencing the variation in frequency.

\section{Evolution of the frequency in circular orbits}\label{Sec5}
For simplicity, we focus on a binary system in a circular orbit and just consider the balance of the angular momentum flux. We can rewrite the velocity in the orthogonal frame $\bm{v}=\dot{r}\bm{n}+r\omega\bm{\tau}$, $\bm{\tau}$ is the unit tangent vector in the orbit plane and $\bm{n}\cdot\bm{\tau}=0$, and $\omega$ is the angular velocity of binary. We can get the motion in this frame as
\begin{equation}\label{orthogonalframe}
    \dot{\bm{n}}=\frac{1}{r}(\bm{v}-\dot{r}\bm{n})=\omega\bm{\tau},\qquad\dot{\bm{\tau}}=-\omega\bm{n},
\end{equation}
and the acceleration is
\begin{align}
    \bm{a}=\bm{\dot{v}}=(\ddot{r}-\omega^2r)\bm{n}+(\dot{\omega}r+2\omega \dot{r})\bm{\tau},\qquad\bm{a}\cdot\bm{n}=\ddot{r}-\omega^2r,\qquad\bm{a}\cdot\bm{v}=\ddot{r}\dot{r}+\omega^2r\dot{r}+\omega\dot{\omega}r^2,
\end{align}
for the tangent direction we get
\begin{align}
    \bm{a}\cdot \bm{\tau}=\dot{\omega}r+2\omega \dot{r}=\Big\{-\frac{GM}{r^2}\Big(\Tilde{\mathcal{B}}_0+\Tilde{\mathcal{B}}_{MT}\frac{\Delta}{\mu}r\chi\Big)+(1+\Tilde{\mathcal{C}}_{MT})\Big\}\omega r.
\end{align}
 The second equality relation comes from Eq.\eqref{CoMacceleration}. Using the relationship between $\dot{r}$ and $\dot{\omega}$ in Appendix \ref{AppE}, we obtain the time derivative of the  angular frequency in quasi-circular orbit 

\begin{align}\label{variationoemga}
    \frac{\dot{\omega}}{\omega}=&-3\bigg\{1+\frac{1}{c^2}\bigg(3+\frac{1}{2}\eta\bigg)(GM)^{2/3}\omega^{2/3}+\frac{1}{c^4} \left(-\frac{1}{3} \eta ^2+\frac{85}{8}  \eta -18 \right)(GM)^{4/3}\omega^{4/3}\bigg\}\frac{\Delta\chi}{M\eta}\notag\\
    &+\frac{1}{c^5}\bigg\{1+\frac{1}{c^2}(GM)^{2/3}\bigg(-\frac{743}{336}-\frac{11}{4}\eta\bigg)\omega^{2/3}\bigg\}\frac{96}{5}(GM)^{5/3}{\eta}\omega^{8/3}.
\end{align}
 
Actually, this equation can also be derived using the conservation of the angular momentum. Turning back to the Lagrangian in Eq.\eqref{CoML} and transforming to the spherical coordinate system $\mathcal{L}(q,\dot{q},\ddot{q})$ with $q=r,\theta,\phi$, choosing the orbit plane $\theta=\pi/2$ and denoting $\omega=\dot{\phi}$, we get the 2PN order orbit angular momentum $L_o$ with MT 
\begin{align}
     L_o&=\frac{\partial \mathcal{L}}{\partial \dot{\phi}}-\frac{\mathrm{d}}{\mathrm{d}t}\frac{\partial \mathcal{L}}{\partial\ddot{\phi}}\\
    &=\mu r^2\dot{\phi}+\frac{1}{c^2}\bigg\{\bigg(\frac{1}{2}-\frac{3}{2}\eta\bigg)(\dot{r}^2+r^2\omega^2)+\frac{GM}{r}(3+\eta)\bigg\}\mu r^2\dot{\phi}\notag\\
    &+\frac{1}{c^4} \bigg\{\frac{G^2M^2}{r^2} \left( \eta^2-\frac{41}{4} \eta+\frac{7}{2}\right)+\frac{GM}{r} \left(-\frac{9}{2} \eta^2 \omega^2 r^2-5 \eta \omega^2 r^2 + 2 \eta^2 \dot{r}^2+4 \eta \dot{r}^2-\frac{7}{2}  \dot{r}^2+\frac{7}{2}r^2 \omega^2 \right)\notag\\
    &\qquad+\frac{39}{8} \eta^2 \omega^4 r^4-\frac{21}{8} \eta \omega^4 r^4+\frac{39}{4} \eta^2 \omega^2 r^2 \dot{r}^2-\frac{21}{4} \eta \omega^2 r^2 \dot{r}^2+\frac{39}{8} \eta^2 \dot{r}^4-\frac{21}{8} \eta \dot{r}^4+\frac{3}{4} \omega^2 r^2 \dot{r}^2+\frac{3}{8} \dot{r}^4+\frac{3}{8} \omega^4 r^4\notag\\
    &\qquad+G \Delta \chi \left(6 \dot{r}-\frac{3 \dot{r}}{4 \eta}\right)\bigg\}+\mathcal{O}(c^{-6}).
\end{align}
Using the balance of the angular momentum flux
\begin{align}
\dot{L}_o+\dot{L}_{GW}=0
\end{align}
and  the relations in the Appendix \ref{AppE}, we can also obtain 
the same result as Eq.\eqref{variationoemga}. Here, ${L}_{GW}$ is the angular momentum of GWs \cite{PhysRevLett.70.113,PhysRevD.52.6882}.

The first line of Eq.\eqref{variationoemga} is the effect of MT at the 0PN, 1PN and 2PN orders, the second line is the effect of GW radiation at 0PN and 1PN orders in quasi-circular orbit. We write the contribution from each term by
\begin{gather}\label{frequencyeach}
    \dot{\omega}_{\text{Newtonian MT}}=-3\frac{\Delta\chi}{M\eta}\omega,\quad \dot{\omega}_{\text{1PN MT}}=-3\frac{1}{c^2}\bigg(3+\frac{1}{2}\eta\bigg)(GM)^{2/3}\frac{\Delta\chi}{M\eta}\omega^{5/3},\\
    \dot{\omega}_{\text{2PN MT}}=-3\frac{1}{c^4} \left(-\frac{1}{3} \eta ^2+\frac{85}{8}  \eta -18 \right)(GM)^{4/3}\frac{\Delta\chi}{M\eta}\omega^{7/3},\\
    \dot{\omega}_{\text{Newtonian GW}}=\frac{1}{c^5}\frac{96}{5}(GM)^{5/3}\omega^{11/3},\quad \dot{\omega}_{\text{1PN GW}}=\frac{1}{c^7}\frac{96}{5}(GM)^{7/3}\bigg(-\frac{743}{336}-\frac{11}{4}\eta\bigg)\omega^{13/3}.
\end{gather}

Here, $\omega$ denotes the total frequency, which includes contributions from all terms, and is obtained by numerically solving the differential equation \eqref{variationoemga}. Substituting $\omega$ back into the equations above allows us to calculate the contribution of each individual term. The frequency of GWs is $f_{GW}=2f_{orbit}=\omega/\pi$. We get the number of GW cycles by the integration of time from $t_1$ to $t_2$ as 
\begin{equation}\label{Ncyc}
\mathcal{N}_{cyc}=\int^{t_2}_{t_1} f_{GW}(t) \mathrm{d}t=\int^{t_2}_{t_1}\mathrm{d}t\bigg[\bigg(\int^{t}_{0}\dot{f}_{GW}(t')\mathrm{d}t'\bigg)+f_0\bigg],
\end{equation}
which can be computed numerically. Similarly, we can get the GW cycles of each term by integrating frequencies in Eq.\eqref{frequencyeach}.

\begin{table*}[htbp!]
  \centering
  \caption{The contribution of MT and GWs to the numbers of cycles $\mathcal{N}_{cyc}$ of GWs. We numerically calculate the effects of each terms in Eq.\eqref{variationoemga} on the number of cycles of GWs after the evolution over a period of 5 years ($t_1=0,t_2=5\text{year}$) of binary systems within quasi-circular orbits. We choose initial conditions involve masses of binary $m_A=1 M_\odot$, $m_B=0.2 M_\odot$, frequency of GWs $f_0=10 \text{mHz},\; 100\text{mHz}$ and the MT rate $\chi=1\times 10^{-5}M_{\odot}/\text{yr},\; 1\times 10^{-6}M_{\odot}/\text{yr}$.}
  \renewcommand\arraystretch{1.8}
    \setlength{\tabcolsep}{12pt}
    \begin{tabular}{c|c|c|c|c}
    \hline
    \hline
    the MT rate $\chi$ & \multicolumn{2}{c|}{$1\times 10^{-5}M_{\odot}/\text{yr}$} &  \multicolumn{2}{c}{$1\times 10^{-6}M_{\odot}/\text{yr}$}    \\
    \hline
    Frequency $f_0 \;(\text{mHz})$ &$100.0 \thicksim 163.0$ & $10\thicksim 10-1.2\times10^{-8}$& $100.0-163.2$& $10\thicksim10+5\times10^{-6}$ \\
    \hline
    \hline
    Total &$\;\;\,1.93\times 10^7$ &$\;\;\,1.58\times 10^6$ & $\;\;\,1.93\times 10^7$&$\,1.58\times 10^6$  \\
    \hline
    Constant frequency $f_0$ &$\;\;\,1.58\times 10^7$ &$\;\;\,1.58\times 10^6$ & $\;\;\,1.58\times 10^7$&$\,1.58\times 10^6$  \\
    \hline
    Total - Newtonian &$-1.56\times 10^{3}$ &$-0.087$ & $-1.56\times 10^{3}$&$-0.045$  \\
    \hline
    Newtonian MT &$-5.25\times 10^3$ & $-4.66\times 10^2$& -525.13&$-46.56$  \\
    \hline
    1PN MT & $-2.68$& $-0.047$& -0.268&$-4.70\times 10^{-3}$  \\
    \hline
    2PN MT &$\;\;\;\;\;2.41\times 10^{-3}$ &$\;\;\;\;\;8.32\times 10^{-6}$  & $\;\;\;\;\;2.41\times 10^{-4}$&$-8.32\times 10^{-7}$  \\
    \hline
    Newtonian GW & $\;\;\,3.53\times 10^{6}$& $\;\;\,4.65\times 10^2$&$\;\;\,3.54\times 10^{6}$ &$\,4.65\times 10^2$  \\
    \hline
    1PN GW& $-1.56\times 10^{3}$&$-0.040$ & $-1.56\times 10^{3}$&$-0.040$  \\
    \hline

    \hline
    \end{tabular}%
    \label{Table}
\end{table*}%

\begin{figure}[ht!]
    \centering
{\includegraphics[width=0.48\textwidth]{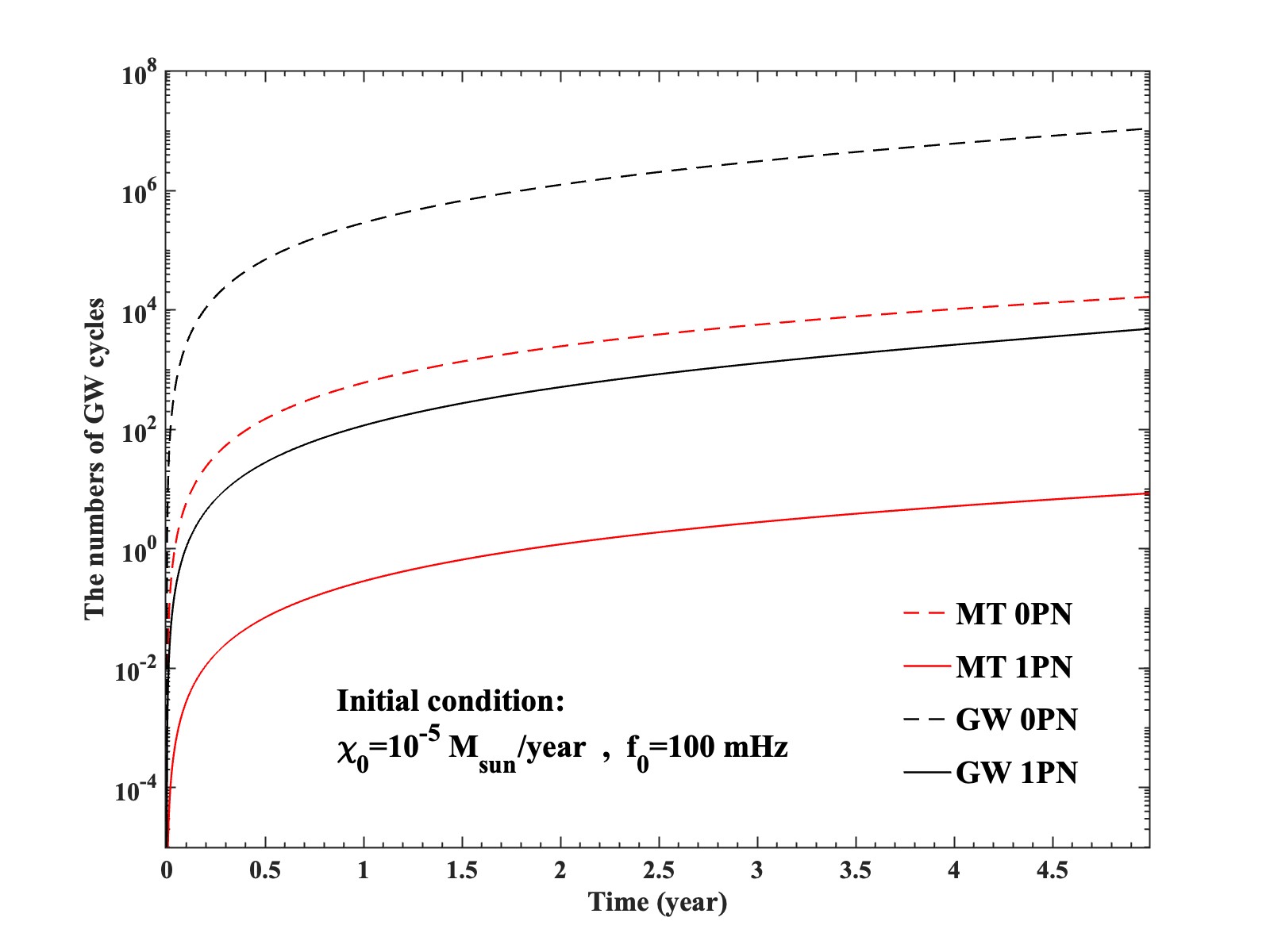}
 \includegraphics[width=0.48\textwidth]{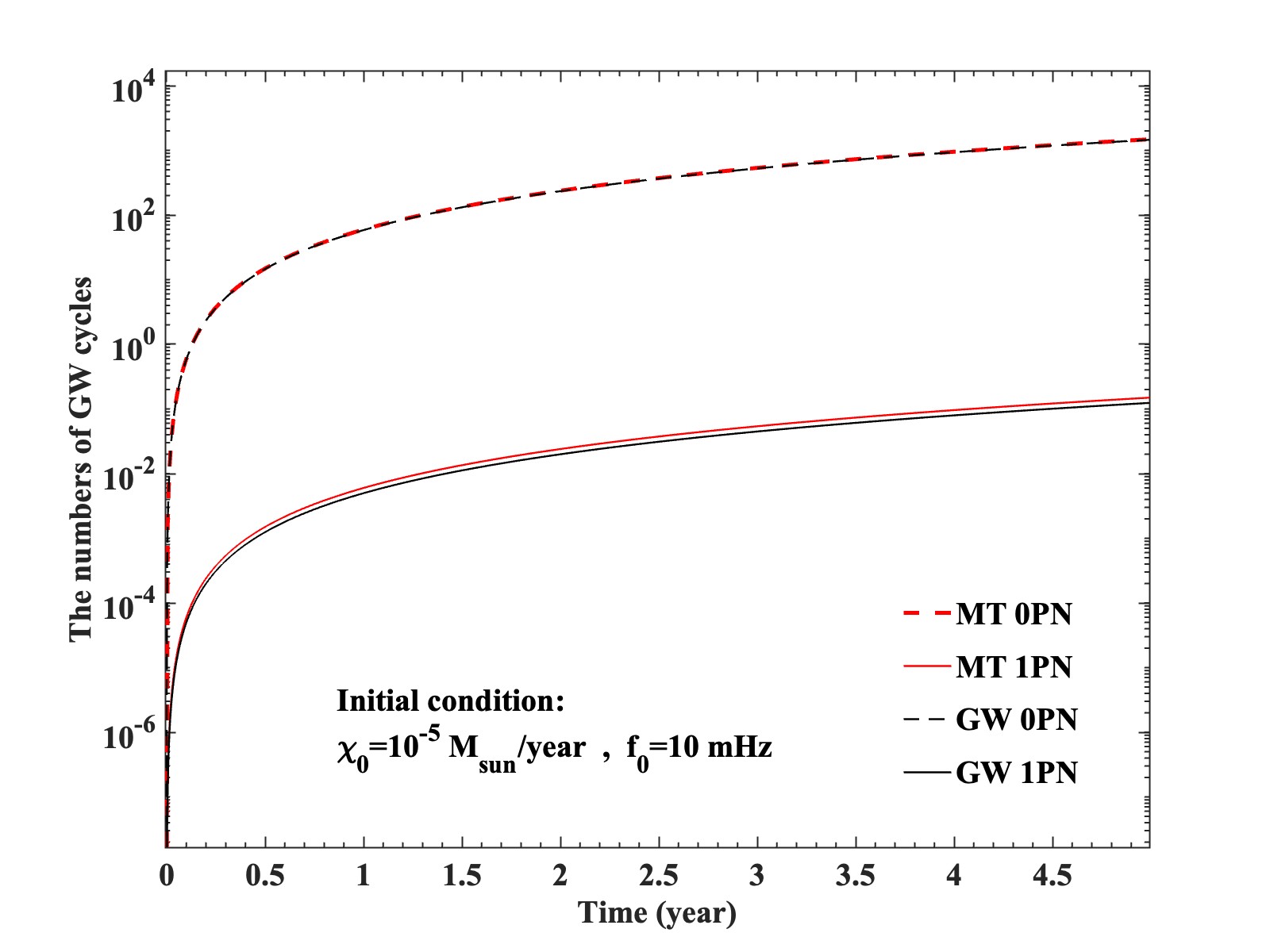}}    
    \caption{The contribution of MT and GWs to the numbers of cycles at the Newtonian-order and the 1PN order. The red lines denote the contribution of MT while the black lines denote the contribution of GWs. The Newtonian order is plotted as dashed lines, and the 1PN order as solid lines. The parameters of the binary system are the same as those of the first column and the second column of Table \ref{Table}. The MT contribution at the Newtonian order is much smaller than the GW contribution at the Newtonian order, but larger than the GW contribution at the 1PN order. Given that the cumulative 1PN order MT contribution to the GW phase shift may exceed 2$\pi$ radians, this effect becomes observable provided the signal-to-noise  is high enough. }
    \label{MTGW}
\end{figure}

In Table \ref{Table}, by relying on  Eq.\eqref{variationoemga}, numerical integration is employed to compute the contribution of each term to the number of cycles of GWs within quasi-circular binary systems after a five-year evolutionary period. The ``Total - Newtonian'' represents the total effect minus the Newtonian effect, which have already been computed in \cite{PhysRevD.111.043049}. At the frequency of 100 mHz, the effect of MT at the 1PN order reaches $-2.68$ cycles, which have a detectable signal on GW detection due to the long-term cumulative results. As shown in Fig. \ref{MTGW}, for a typical binary system, the MT contribution at Newtonian order is much smaller than the GW contribution at Newtonian order, but larger than the GW contribution at 1PN order. 

Regarding the impact of masses of binaries, it can be seen from the characteristic term $\Delta\chi/(M\eta)$ of the MT effect in Eq.\eqref{variationoemga} that the MT effect is inversely proportional to the mass of the companion star $\Delta\chi/(M\eta)\propto m_B^{-1}$. For companion stars with smaller masses, the MT effect would be more significant.

\section{Conclusion}
We present a derivation of the 2.5PN equations of motion of compact binaries with MT. Applying the result to compact binaries in quasi-circular orbits, we find that the MT effect at the 1PN order is non-negligible.
The 2.5PN acceleration with the MT effect in Eq.\eqref{CoMacceleration} also lays the groundwork for future studies of binaries in the case of elliptical orbits and the higher the MT rates.


It is worth noting that in this work, we do not impose constraints or explicit expressions for the evolution of the MT rate $\dot{m}$, allowing the MT rate to evolve in an arbitrary form, including the results of numerical fitting or astronomical observations. This flexibility enables the formulation to accommodate various the MT rates. However, if $\dot{m}$ strongly depends on the binary separation or velocity, and if accurate analytical expressions can be provided, a more rigorous analysis of $\partial \dot{m} / \partial v$ and $\partial \dot{m} / \partial r$ in the Euler-Lagrange equation should be undertaken.

The acceleration with MT corrections can help extract more information about binary systems from GW signals, including the individual masses of the binaries, their equations of state, and other important properties. Additionally, it can improve the accuracy of parameter estimation and explain the observational phenomena of binary systems exhibiting increasing orbital periods. Typically, stronger relativistic effects and the MT effect occur during the late inspiral phase, close to the merger, which is better calculated by numerical relativity. However, in the stable accretion process discussed in this work, the 2.5PN  approximation provides both sufficient precision and computational efficiency. This makes it particularly useful for analyzing GW signals during the long-term inspiral phase, offering advantages for extended-duration signal analysis.

\section{ACKNOWLEDGMENTS}
T. L. is supported by the China Postdoctoral Science Foundation Grant No. 2024M760692. This work is supported in part by the National Key Research and Development Program of China Grant No. 2020YFC2201501, in part by the National Natural Science Foundation of China under Grant No. 12475067 and No. 12235019.

\appendix

\section{Calculation of Post-Newton potential}\label{A}

In this section we will calculate the post-Newton potential in Eq.\eqref{VU} and Eq.\eqref{UT}. Firstly for $\partial^2_t \Delta^{-1}U$ term of order order $(c^{-2})$ so we need to expand $T^{\mu\nu}$ to $(c^{-2})$ order and get the derivative of mass with respect to time as $\dot{m}$ and $\ddot{m}$, then we get
\begin{align}\label{Vtm}
      \partial^2_t 2\Delta^{-1}U&=\partial^2_t 2\Delta^{-2}(-4\pi G \sigma)=\frac{G}{c^2}\sum_B\partial_t^2\int\mathrm{d}^3\bm{x}'|\bm{x}-\bm{x}'|\left[T^{00}+T^{ii}\right]_B(\bm{x}',t)\notag\\
      &=\sum_{B}\partial_t^2Gm_B\int\mathrm{d}^3\bm{x}'|\bm{x}-\bm{x}'|\delta^3(\bm{x}-\bm{x}_B)\left\{1+\frac{1}{c^2}\left[\frac{1}{2}v_A^2-V_A+v_A^iv_A^i\right]+\mathcal{O}(c^{-4})\right\}\notag\\
      &=\sum_{B}G\bigg\{\ddot{m}_B r_{B}+2\dot{m}_B\dot{ r}_{B}+m_B\ddot{r}_{B}\notag\\
      &+\frac{1}{c^2}\bigg\{m_B\Big(6 a_B v_B \dot{r}_B+3 r_B a_B^2 +\frac{3}{2} v_B^2 \ddot{r}_B+3 r_B v_B \dot{a}_B\Big)+\dot{m}_B \left(6 r_B a_B v_B+3 v_B^2 \dot{r}_B\right)+\frac{3}{2} r_B v_B^2 \ddot{m}_B\notag\\
      &\qquad+\sum_{C\neq B} \frac{G}{r_{BC}} \bigg[2 r_B \dot{m}_B \dot{m}_C+2 m_C \dot{m}_B\dot{r}_B+2 m_B \dot{r}_B \dot{m}_C+r_B m_C \ddot{m}_B+m_B r_B \ddot{m}_C+m_B m_C \ddot{r}_B\notag\\
      &\qquad\qquad +r_B m_B  m_C \Big(2\frac{\dot{r}_{BC}^2}{r_{BC}^2}- \frac{\ddot{r}_{BC}}{r_{BC}}\Big)-\frac{\dot{r}_{BC}}{r_{BC}}\Big(2r_B m_B  \dot{m}_C+2 m_B m_C \dot{r}_B+2 r_B m_C \dot{m}_B\Big) \bigg]\bigg\}\bigg\}+\mathcal{O}(c^{-4})\notag\\
      &=\sum_B G\bigg\{{\ddot{m}_Br_{B}+2\dot{m}_B(\bm{v}_{B}\cdot\bm{n}_{B})}+m_B\Big(\bm{a}_{B}\cdot\bm{n}_{B}-\frac{1}{r_B}(\bm{v}_{B}\cdot\bm{n}_{B})^2+\frac{1}{r_{B}}v^2_{B}\Big)\notag\\
      &+\frac{1}{c^2}\bigg\{m_B\bigg[6 a_B v_B (\bm{v}_{B}\cdot\bm{n}_{B})+3 r_B a_B^2 +\frac{3}{2} v_B^2 \Big(\bm{a}_{B}\cdot\bm{n}_{B}-\frac{1}{r_B}(\bm{v}_{B}\cdot\bm{n}_{B})^2+\frac{1}{r_{B}}v^2_{B}\Big)\bigg]+\frac{3}{2} r_B v_B^2 \ddot{m}_B\notag\\
      &\qquad+\dot{m}_B \Big(6 r_B a_B v_B+3 v_B^2 (\bm{v}_{B}\cdot\bm{n}_{B})\Big)-3r_B{\dot{m}_B}(\bm{a}_B\cdot\bm{v}_B)-3r_B{\ddot{m}_B}{v}_B^2+r_B\frac{\dot{m}_B^2}{m_B}{v}_B^2\notag\\
      &\qquad+\sum_{C\neq B} \frac{G}{r_{BC}} \bigg[2 r_B \dot{m}_B \dot{m}_C+2 m_C \dot{m}_B(\bm{v}_{B}\cdot\bm{n}_{B})+2 m_B (\bm{v}_{B}\cdot\bm{n}_{B}) \dot{m}_C+r_B m_C \ddot{m}_B+m_B r_B \ddot{m}_C\notag\\
      &\qquad\qquad+m_B m_C \Big(\bm{a}_{B}\cdot\bm{n}_{B}-\frac{1}{r_B}(\bm{v}_{B}\cdot\bm{n}_{B})^2+\frac{1}{r_{B}}v^2_{B}\Big)-3\frac{r_B}{r_{BC}}m_B\dot{m}_C(\bm{v}_B\cdot\bm{n}_{BC})\notag\\
      &\qquad\qquad +r_B m_B  m_C \bigg[\frac{2}{r_{BC}^2}(\bm{v}_{BC}\cdot\bm{n}_{BC})^2- \frac{1}{r_{BC}}\Big(\bm{a}_{B}\cdot\bm{n}_{B}-\frac{1}{r_{BC}}(\bm{v}_{BC}\cdot\bm{n}_{BC})^2+\frac{1}{r_{BC}}v^2_{BC}\Big)\notag\\
      &\qquad\qquad+\frac{6}{r_{BC}^2}(\bm{v}_{BC}\cdot\bm{n}_{BC})(\bm{v}_B\cdot\bm{n}_{BC})-\frac{3}{r_{BC}^2}(\bm{v}_B\cdot\bm{v}_{BC})\bigg]\notag\\
      &\qquad\qquad-\frac{1}{r_{BC}}(\bm{v}_{BC}\cdot\bm{n}_{BC})\Big(2r_B m_B  \dot{m}_C+2 m_B m_C (\bm{v}_{B}\cdot\bm{n}_{B})+2 r_B m_C \dot{m}_B\Big) \bigg]\bigg\}\bigg\}+\mathcal{O}(c^{-4}).
\end{align}
For $\partial^4_t \Delta^{-2}U$ term with $(c^{-4})$ order $T^{\mu\nu}$ can only remain the leading order, while $U=\Delta^{-1}(-4\pi G\sigma)$ there are totally $\Delta^{-3}$ and use the Poison integration in Eq.\eqref{PoisonI} then
\begin{align}
     \partial^4_t 24\Delta^{-2}U&=\partial^4_t 24\Delta^{-3}(-4\pi G \sigma)=\frac{G}{c^2}\sum_B\partial_t^2\int\mathrm{d}^3\bm{x}'|\bm{x}-\bm{x}'|^3\left[T^{00}+T^{ii}\right]_B(\bm{x}',t)\notag\\
     &=\sum_Bm_B\bigg\{12 r_B^2 {\mathop{m_B}^{(3)}} \dot{r}_B+3 r_B m_B \left(r_B {\mathop{r_B}^{(4)}}+6 \ddot{r}_B^2\right)+24 {\mathop{r_B}^{(3)}} r_B m_B \dot{r}_B+36 m_B \dot{r}_B^2 \ddot{r}_B+\ddot{m}_B \left(18 r_B^2 \ddot{r}_B+36 r_B \dot{r}_B^2\right)\notag\\
      &\qquad+\dot{m}_B \left(12 r_B^2 {\mathop{r_B}^{(3)}}+24 \dot{r}_B^3+72 r_B \dot{r}_B \ddot{r}_B\right)+r_B^3 {\mathop{m_B}^{(4)}}\bigg\}+\mathcal{O}(c^{-2}),
\end{align}
where $|\bm{x}_B|=r_{B}$ and $\bm{r}_B=r_B\bm{n}_B$, the derivative of $r_B$ with time can be calculated as
\begin{align}
    \dot{r}_B&=\bm{v}_B\cdot\bm{n}_B,\\
    \ddot{r}_B&=\bm{a}_B\cdot\bm{n}_B+\frac{1}{r_B}v_B^2-\frac{1}{r_B}(\bm{v}_B\cdot\bm{n}_B)^2,\\
    \mathop{r_B}^{(3)}&=\frac{3}{r_B^2}\Big[ (\bm{a}_B\cdot\bm{v}_B)r_B-( \bm{v}_B\cdot\bm{n}_B)\big((\bm{a}_B\cdot\bm{n}_B)r_B- v_B^2+ (\bm{v}_B\cdot\bm{n}_B)^2\big)\Big]+\sum_C\bigg\{\frac{\dot{m}_C}{m_C} \left(-\frac{G m_C}{r_{BC}^2}\bm{n}_B\cdot \bm{n}_{BC}-\bm{a}_B\cdot\bm{n}_B\right)\notag\\
    &+\frac{3 Gm_C}{r_{BC}^3} \bm{n}_B\cdot\bm{n}_{BC} (\bm{v}_{BC}\cdot\bm{n}_{BC})-\frac{Gm_C}{r_{BC}^3} \bm{n}_B\cdot\bm{v}_{BC} +\frac{\dot{m}_C^2}{m_C^2}\bm{v}_B\cdot\bm{n}_B -\frac{\ddot{m}_C}{m_C}\bm{v}_B\cdot\bm{n}_B \bigg\},\\
    \mathop{r_B}^{(4)}&=-\frac{6 }{r_B^2}\Big[2 (\bm{a}_B\cdot\bm{v}_B) (\bm{v}_B\cdot\bm{n}_B)+(\bm{a}_B\cdot\bm{n}_B) \left(v_B^2-3 (\bm{v}_B\cdot\bm{n}_B)^2\right)\Big]-\frac{3}{r_B^3} \Big[-6 (\bm{v}_B\cdot\bm{n}_B)^2 v_B^2+5 (\bm{v}_B\cdot\bm{n}_B)^4+v_B^4\Big]\notag\\
    &+\sum_C\bigg\{\frac{G m_C}{r_{BC}^4}\bigg[-r_{BC} \left(\bm{a}_{BC}-3 \bm{n}_{BC} (\bm{a}_{BC}\cdot\bm{n}_{BC})\right)+3 \bm{n}_{BC} \left(v_{BC}^2-5 (\bm{v}_{BC}\cdot\bm{n}_{BC})^2\right)+6 (\bm{v}_{BC}\cdot\bm{n}_{BC}) v_{BC}\bigg]\bm{n}_B\notag\\
    &\qquad+\frac{3}{r_B}\left(a_{B}^2-(\bm{a}_B\cdot\bm{n}_B)^2\right)-4 \frac{G m_C}{r_Br_{AB}^3} \left(\bm{n}_B (\bm{v}_B\cdot\bm{n}_B)-\bm{v}_B\right) \left(3 \bm{n}_{BC} (\bm{v}_{BC}\cdot\bm{n}_{BC})-\bm{v}_{BC}\right)\notag\\
    &\qquad+\frac{\dot{m}_C}{m_C}\bigg[4 \frac{Gm_C}{r_Br_{BC}^2}\bm{n}_{BC}  \left(\bm{n}_B (\bm{v}_B\cdot\bm{n}_B)-\bm{v}_B\right)-\frac{G m_C}{r_{BC}^3} 
     \left(\bm{v}_{BC}-3 \bm{n}_{BC} (\bm{v}_{BC}\cdot\bm{n}_{BC})\right)\cdot\bm{n}_B\notag\\
    &\qquad\qquad\quad+\frac{4}{r_B} \left((\bm{a}_B\cdot\bm{n}_B) (\bm{v}_B\cdot\bm{n}_B)-(\bm{a}_B\cdot\bm{v}_B)\right)\bigg]-(\bm{v}_B\cdot\bm{n}_B) \frac{\dddot{m}_C}{m_C}-3 (\bm{v}_B\cdot\bm{n}_B) \frac{\dot{m}_C^3}{m_C^3}+4 (\bm{v}_B\cdot\bm{n}_B) \frac{\dot{m}_C\ddot{m}_C}{m_C^2}\notag\\
    &\qquad+\frac{\dot{m}_C^2}{m_C^2}\bigg[\frac{G m_C }{r_{BC}^2}(\bm{n}_B\cdot\bm{n}_{BC})+\frac{4}{r_B}\left(v_B^2-(\bm{v}_B\cdot\bm{n}_B)^2\right) +3 (\bm{a}_B\cdot\bm{n}_B) \bigg]\notag\\
    &\qquad+\frac{\ddot{m}_C}{m_C}\bigg[\frac{G m_C }{r_{BC}^2}(\bm{n}_B\cdot\bm{n}_{BC})+\frac{4}{r_B}\left((\bm{v}_B\cdot\bm{n}_B)^2-v_B^2\right)-2 (\bm{a}_B\cdot\bm{n}_B)\bigg]\bigg\},
\end{align}
we will get rid of the acceleration term $\bm{a}_B$ in 1PN order by the integration by part 
\begin{equation}
    \partial_t(\bm{v}_B\cdot \bm{n}_{AB})=\bm{a}_B\cdot\bm{n}_{AB}+\frac{1}{r_{AB}}\left[\bm{v}_B\cdot\bm{v}_{AB}-(\bm{v}_B\cdot\bm{n}_{AB})(\bm{v}_{AB}\cdot\bm{n}_{AB})\right],
\end{equation}
and we need to note that mass $m_{B}$ is function of $t$, so the integration by part should consider the effect of mass as
\begin{equation}
\partial_t(m_B(\bm{v}_B\cdot\bm{n}_{AB}))=\dot{m}_B(\bm{v}_{B}\cdot\bm{n}_{AB})+m_B\partial_t(\bm{v}_B\cdot\bm{n}_{AB})
\end{equation}
The term with $\dot{a}$ in the highest order $(c^{-4})$ can be calculated by the Newton acceleration with MT as
\begin{gather}
    \bm{a}_B=-\sum_C\frac{G m_C}{r_{BC}^2}\bm{n}_{BC}-\frac{\dot{m}_B}{m_B}\bm{v}_B,\\
    \bm{\dot{a}}_B=-\sum_C\frac{G m_C}{r_{BC}^2}\bigg[\frac{\dot{m}_C}{m_C}\bm{n}_{BC}-\frac{2}{r_{BC}}(\bm{v}_{BC}\cdot\bm{n}_{BC})\bm{n}_{BC}+\frac{1}{r_{BC}}\bm{v}_{BC}\bigg]-\frac{\dot{m}_B}{m_B}\bm{a}_B-\frac{\ddot{m}_B}{m_B}\bm{v}_B+\frac{\dot{m}_B^2}{m_B^2}\bm{v}_B,
\end{gather}

Based on these energy-momentum tensor we can get the expression of $V$ and $U$ with Eq.\eqref{VU} and Eq.\eqref{UT}, it is important that we need  
\begin{align}
      V_A&=\sum_{B\neq A}\frac{Gm_B}{r_{AB}}\Bigg\{1+\frac{1}{c^2}\bigg\{\frac{3v_A^2}{2}-\sum_{C\neq B}\frac{Gm_C}{r_{BC}}+\frac{1}{2}r_{AB}\left[\bm{a}_{B}\cdot\bm{n}_{BA}-\frac{(\bm{v}_{B}\cdot\bm{n}_{BA})^2}{r_{AB}}+\frac{v^2_{B}}{r_{AB}}\right]\notag\\
      &\qquad\qquad\qquad\qquad\qquad+r_{AB}\frac{\dot{m}_B}{m_B}(\bm{v}_B\cdot\bm{n}_{BA})+\frac{1}{2}\frac{\ddot{m}_B}{m_B}r_{AB}^2\bigg\}\notag\\
      &+\frac{1}{c^4}\bigg\{\sum_{C\neq B}\frac{G m_C}{r_{BC}}\bigg[-\frac{11}{2}v_C^2-4\bm{v}_B\bm{v}_C+\frac{1}{2}\Big(\bm{a}_{C}\bm{r}_{BC}+v_B^2-v_C^2+(\bm{v}_C\cdot\bm{n}_{BC})^2\Big)+3\sum_{D\neq C}\frac{Gm_D}{r_{CD}}+\frac{11}{2}\sum_{D\neq B}\frac{Gm_D}{r_{BD}}\notag\\
      &\qquad\qquad+ r_{AB}^2 \frac{\dot{m}_B}{m_B} \frac{\dot{m}_C}{m_C} +r_{AB}\frac{\dot{m}_B}{m_B}(\bm{v}_{B}\cdot\bm{n}_{BA})+ r_{AB} (\bm{v}_{B}\cdot\bm{n}_{BA}) \frac{\dot{m}_C}{m_C}+\frac{1}{2}r_{AB}^2 \frac{\ddot{m}_B}{m_B}+\frac{1}{2} r_{AB}^2 \frac{\ddot{m}_C}{m_C}\notag\\
      &\qquad\qquad+\frac{1}{2} \Big(r_{AB}\bm{a}_{B}\cdot\bm{n}_{BA}-(\bm{v}_{B}\cdot\bm{n}_{BA})^2+v^2_{B}\Big)-\frac{3}{2}\frac{r_{AB}^2}{r_{BC}}\frac{\dot{m}_C}{m_C}(\bm{v}_B\cdot\bm{n}_{BC})\notag\\
      &\qquad\qquad +r_{AB}^2\bigg(\frac{1}{r_{BC}^2}(\bm{v}_{BC}\cdot\bm{n}_{BC})^2-\frac{1}{2} \frac{1}{r_{BC}}\Big(\bm{a}_{B}\cdot\bm{n}_{BA}-\frac{1}{r_{BC}^2}(\bm{v}_{BC}\cdot\bm{n}_{BC})^2+\frac{1}{r_{BC}^2}v^2_{BC}\Big)\notag\\
      &\qquad\qquad\qquad\qquad\quad+\frac{3}{r_{BC}}(\bm{v}_{BC}\cdot\bm{n}_{BC})(\bm{v}_B\cdot\bm{n}_{BC})-\frac{3}{2}\frac{1}{r_{BC}}(\bm{v}_B\cdot\bm{v}_{BC})\bigg)\notag\\
      &\qquad\qquad-\frac{r_{AB}}{r_{BC}}(\bm{v}_{BC}\cdot\bm{n}_{BC})\Big(r_{AB}  \frac{\dot{m}_C}{m_C}+  (\bm{v}_{B}\cdot\bm{n}_{BA})+ r_{AB} \frac{\dot{m}_B}{m_B}\Big)\bigg]\notag\\
      &\qquad+\frac{7}{8}v_B^4+3 r_{AB}(\bm{a}_B \cdot\bm{v}_B) (\bm{v}_{B}\cdot\bm{n}_{B})+\frac{3}{2} r_{AB}^2 a_B^2 +\frac{3}{4} v_B^2 \Big(r_{AB}\bm{a}_{B}\cdot\bm{n}_{BA}-(\bm{v}_{B}\cdot\bm{n}_{B})^2+v^2_{B}\Big)\notag\\
      &\qquad+\frac{3}{2}r_{AB}\frac{\dot{m}_B}{m_B} \Big( r_{AB}(\bm{a}_B\cdot\bm{v}_B)+ v_B^2 (\bm{v}_{B}\cdot\bm{n}_{B})\Big)+r_{AB}^2{v}_B^2\Big(\frac{1}{2}\frac{\dot{m}_B^2}{m_B^2}-\frac{3}{4}\frac{\ddot{m}_B}{m_B}\Big)\notag\\
      &\qquad+r_{AB} m_B \Big(\frac{1}{8}r_{AB} {\mathop{r_{AB}}^{(4)}}+\frac{3}{4} \ddot{r}_{AB}^2\Big)+ {\mathop{r_{AB}}^{(3)}} r_{AB} m_B \dot{r}_{AB}+\frac{3}{2} m_B \dot{r}_{AB}^2 \ddot{r}_{AB}+\ddot{m}_B \Big(\frac{3}{4} r_{AB}^2 \ddot{r}_{AB}+\frac{3}{2} r_{AB} \dot{r}_{AB}^2\Big)\notag\\
      &\qquad\qquad+\dot{m}_B \Big(\frac{1}{2} r_{AB}^2 {\mathop{r_{AB}}^{(3)}}+ \dot{r}_{AB}^3+3 r_{AB} \dot{r}_{AB} \ddot{r}_{AB}\Big)+\frac{1}{2} r_B^2 {\mathop{m_B}^{(3)}} \dot{r}_{AB}+\frac{1}{24}r_{AB}^3 {\mathop{m_B}^{(4)}}\bigg\}\Bigg\}+\mathcal{O}(c^{-6})\notag\\    
      &=\sum_{B\neq A}\frac{Gm_B}{r_{AB}}\Bigg\{1+\frac{1}{c^2}\bigg\{\frac{3v_A^2}{2}-\sum_{C\neq B}\frac{Gm_C}{r_{BC}}+\frac{1}{2}r_{AB}\left[\bm{a}_{B}\cdot\bm{n}_{BA}-\frac{(\bm{v}_{B}\cdot\bm{n}_{BA})^2}{r_{AB}}+\frac{v^2_{B}}{r_{AB}}\right]\notag\\
      &\qquad\qquad\qquad\qquad\qquad+r_{AB}\frac{\dot{m}_B}{m_B}(\bm{v}_B\cdot\bm{n}_{BA})+\frac{1}{2}\frac{\ddot{m}_B}{m_B}r_{AB}^2\bigg\}\notag\\
      &+\frac{1}{c^4}\bigg\{\sum_{C\neq B}\frac{G m_C}{r_{BC}}\bigg[ r_{AB}^2 \frac{\dot{m}_B}{m_B} \frac{\dot{m}_C}{m_C} +r_{AB}(\bm{v}_{B}\cdot\bm{n}_{BA})\frac{\dot{m}_B}{m_B}+ r_{AB} (\bm{v}_{B}\cdot\bm{n}_{BA}) \frac{\dot{m}_C}{m_C}+\frac{1}{2}r_{AB}^2 \frac{\ddot{m}_B}{m_B}+\frac{1}{2} r_{AB}^2 \frac{\ddot{m}_C}{m_C}\notag\\
      &\qquad\qquad-\frac{3}{2}\frac{r_{AB}^2}{r_{BC}}\frac{\dot{m}_C}{m_C}(\bm{v}_B\cdot\bm{n}_{BC})-\frac{r_{AB}^2}{r_{BC}}(\bm{v}_{BC}\cdot\bm{n}_{BC})\Big(  \frac{\dot{m}_C}{m_C}+ \frac{\dot{m}_B}{m_B}\Big)\notag\\
      &\qquad\qquad-\frac{\dot{m}_C}{m_C} \bigg(-\frac{3}{4}\frac{r_{AB}^3}{r_{BC}^2} (\bm{n}_{BA} \cdot\bm{n}_{BC}) (\bm{v}_{BC}\cdot\bm{n}_{BC}) +\frac{1}{4} \frac{r_{AB}^3}{r_{BC}^2}(\bm{v}_{BC}\cdot\bm{n}_{BA}) +\frac{1}{2} \frac{r_{AB}^2}{r_{BC}}(\bm{n}_{BA} \cdot\bm{n}_{BC})(\bm{v}_B\cdot\bm{n}_{BA}) \notag\\
      &\qquad\qquad+\frac{1}{2} \frac{r_{AB}^2}{r_{BC}}(\bm{v}_B \cdot\bm{n}_{BC})\bigg)+\frac{\dot{m}_B}{m_B} \left(- \frac{3}{8} \frac{r_{AB}^3}{r_{BC}^2} (\bm{v}_{BC}\cdot\bm{n}_{BA}) +\frac{9}{8} \frac{r_{AB}^3}{ r_{BC}^2} (\bm{n}_{BA}\cdot\bm{n}_{BC})(\bm{v}_{BC}\cdot\bm{n}_{BC})\right)\notag\\
     &\qquad\qquad-\frac{3}{8} \frac{r_{AB}^3}{ r_{BC}} (\bm{n}_{BA}\cdot\bm{n}_{BC})\frac{\dot{m}_B}{m_B} \frac{\dot{m}_C}{m_C}-\frac{1}{8} \frac{r_{AB}^3}{ r_{BC}} (\bm{n}_{BA}\cdot \bm{n}_{BC})\frac{\ddot{m}_C}{m_C}\bigg]\notag\\
      &\qquad+\frac{\dot{m}_B}{m_B} \left(\frac{5}{2}r_{AB}^2 (\bm{a}_B\cdot\bm{v}_B)+r_{AB}^2 (\bm{a}_B\cdot\bm{n}_{BA}) (\bm{v}_B\cdot\bm{n}_{BA})+3 r_{AB} (\bm{v}_B\cdot\bm{n}_{BA}) v_B^2-\frac{1}{2} r_{AB} (\bm{v}_B\cdot\bm{n}_{BA})^3\right)\notag\\
      &\qquad+\frac{\ddot{m}_B}{m_B} \left(\frac{1}{2} r_{AB}^3 (\bm{a}_B\cdot\bm{n}_B)+\frac{1}{4} r_{AB}^2 (\bm{v}_B\cdot\bm{n}_{BA})^2-\frac{1}{2} r_{AB}^2 v_B^2\right)\notag\\
      &\qquad+\frac{\dot{m}_B^2}{m_B^2} \left(-\frac{1}{8} r_{AB}^3 (\bm{a}_B\cdot\bm{n}_B)+\frac{1}{2} r_{AB}^2 (\bm{v}_B\cdot\bm{n}_{BA})^2+ r_{AB}^2 v_B^2\right)\bigg\}\Bigg\}\nonumber\\
      &+V_{A(\text{2PN without MT})}+\mathcal{O}(c^{-6}).
\end{align}
and $V_A^i$ can be written as
\begin{align}
V_{A}^i=&\sum_{B\neq A}\frac{Gm_B}{r_{AB}}\left\{v_B^i\left[1+\frac{1}{c^2}\left(\frac{1}{2}v_B^2-\sum_{C\neq B}\frac{Gm_C}{r_{BC}}\right)\right]+{\frac{1}{c^2}\frac{1}{2}\left[\frac{r_{AB}}{m_B}\partial_t^2\left(r_{AB}m_Bv_B^i\right)\right]}\right\}\nonumber\\
=&\sum_{B\neq A}\frac{Gm_B}{r_{AB}}\Bigg\{v_B^i+{\frac{1}{c^2}\left[\frac{1}{2}\frac{\ddot{m}_B}{m_B}r_{AB}^2v_B^i+\frac{\dot{m}_B}{m_B}\left(r_{AB}(\bm{v}_{B}\cdot\bm{n}_{BA})v_B^i+r_{AB}^2a_B^i\right)\right]}\Bigg\}\nonumber\\
&+V^i_{A(\text{2PN without MT})}+\mathcal{O}(c^{-4}).
\end{align}
where $V_{A(\text{2PN without MT})}$ is the normal 2PN terms without MT. In the following calculations we can ignore their product with MT terms, because the effect of these terms will appear in the order of 3PN. For potential $V$ the effects of $\dot{m}$ appear in 1PN first and do not appear in Newton order because of the existing of $c^{-2}\partial_t^2$. In addition, we need to note that the subscript $A$ of $V$ means the potential at point $A$ caused by $B$ or other points, so the translation form the Newton potential $U$ at origin point of coordinates to post-Newton potential at point A can replace $\bm{n}_B$ and $r_B$ by $\bm{n}_{BA}$ and $r_{AB}$, but keep $\bm{v}_B$ and $\bm{a}_B$ invariant.

\section{2PN Lagrangian of the N-body system}\label{AppB}
Using the post-Newtonian potential $V$ in Appendix \ref{A} and the 2PN Fokker action in Eq.\eqref{Action} to get the Lagrangian of the N-body system as following
\begin{align}
    \mathcal{L}=&\sum_{A}m_A\Bigg\{\frac{v_A^2}{2}+\frac{1}{2}\sum_{B\neq A}\frac{Gm_B}{r_{AB}}+\frac{1}{c^2}\Bigg\{\frac{v_A^4}{8}+\sum_{B\neq A}\frac{Gm_B}{r_{AB}}\bigg[\frac{3}{4}v_A^2+\frac{3}{4}v_B^2-\frac{7}{4}\bm{v}_A\cdot\bm{v}_B-\frac{1}{4}(\bm{v}_A\cdot\bm{n}_{AB})(\bm{v}_B\cdot\bm{n}_{AB})\notag\\
    &\qquad\qquad\qquad\qquad\qquad\qquad\qquad\quad-\frac{1}{2}\sum_{C\neq B}\frac{Gm_C}{r_{BC}}+\frac{1}{4}\frac{\dot{m}_B}{m_B}(\bm{v}_B\cdot\bm{n}_{BA})r_{AB}+\frac{1}{4}\frac{\ddot{m}_B}{m_B}r_{AB}^2\bigg]\Bigg\}\notag\\
    &+\frac{1}{c^4}\Bigg\{\frac{v_A^6}{16}+\sum_{B\neq A}\frac{Gm_B}{r_{AB}}\Bigg[\sum_{C\neq B}\frac{G^2m_C^2}{2r_{BC}^2}+\sum_{C\neq B}\frac{19G^2m_Bm_C}{8r_{BC}r_{AB}}+\frac{3}{16}(\bm{v}_A\cdot\bm{n}_{AB})^2(\bm{v}_B\cdot\bm{n}_{AB})^2-\frac{7}{8}(\bm{v}_B\cdot\bm{n}_{AB})^2v_A^2+\frac{7}{8}v_A^4\notag\\
    &\qquad-2v_A^2(\bm{v}_A\cdot\bm{v}_{B})+\frac{1}{8}(\bm{v}_A\cdot\bm{v}_{B})^2+\frac{3}{4}(\bm{v}_A\cdot\bm{n}_{AB})(\bm{v}_B\cdot\bm{n}_{AB})(\bm{v}_A\cdot\bm{v}_{B})+\frac{15}{16}v_A^2v_B^2\notag\\
    &\qquad+r_{AB}\Big(-\frac{7}{4}(\bm{a}_A\cdot\bm{v}_{B})(\bm{v}_B\cdot\bm{n}_{AB})-\frac{1}{8}(\bm{a}_A\cdot\bm{n}_{AB})(\bm{v}_B\cdot\bm{n}_{AB})^2+\frac{7}{8}(\bm{a}_A\cdot\bm{n}_{AB})v_B^2\Big)\notag\\
    &\qquad+\sum_{C\neq B}\frac{Gm_C}{r_{BC}}\bigg(\frac{7}{2}(\bm{v}_C\cdot\bm{n}_{CB})^2-\frac{7}{2}(\bm{v}_B\cdot\bm{n}_{CB})(\bm{v}_C\cdot\bm{n}_{CB})+\frac{1}{2}(\bm{v}_B\cdot\bm{n}_{CB})^2+\frac{1}{4}v_C^2-\frac{7}{4}(\bm{v}_C\cdot\bm{v}_{B})+\frac{7}{4}v_B^2\bigg)\Bigg]\Bigg\}\notag\\
    &+\frac{1}{c^4}\sum_{B\neq A}\frac{Gm_B}{r_{AB}}\Bigg\{\sum_{D\neq A}\frac{G m_D}{r_{AD}}\bigg[-\frac{1}{4}r_{AD}^2\frac{\ddot{m}_D}{m_D}-\frac{1}{2}r_{AD}\frac{\dot{m}_D}{m_D}(\bm{v}_D\cdot\bm{n}_{DA})\bigg]\notag\\
    &\qquad+\sum_{C\neq B}\frac{G m_C}{r_{BC}}\Bigg[ \frac{1}{2}r_{AB}^2 \frac{\dot{m}_B}{m_B} \frac{\dot{m}_C}{m_C} +\frac{1}{2}r_{AB}(\bm{v}_{B}\cdot\bm{n}_{BA})\frac{\dot{m}_B}{m_B}+ \frac{1}{2}r_{AB} (\bm{v}_{B}\cdot\bm{n}_{BA}) \frac{\dot{m}_C}{m_C}+\frac{1}{4}r_{AB}^2 \frac{\ddot{m}_B}{m_B}+\frac{1}{4} r_{AB}^2 \frac{\ddot{m}_C}{m_C}\notag\\
      &\qquad\qquad-\frac{3}{4}\frac{r_{AB}^2}{r_{BC}}\frac{\dot{m}_C}{m_C}(\bm{v}_B\cdot\bm{n}_{BC})-\frac{1}{2}\frac{r_{AB}^2}{r_{BC}}(\bm{v}_{BC}\cdot\bm{n}_{BC})\Big(  \frac{\dot{m}_C}{m_C}+ \frac{\dot{m}_B}{m_B}\Big)\notag\\
      &\qquad\qquad-\frac{\dot{m}_C}{m_C} \bigg(-\frac{3}{8}\frac{r_{AB}^3}{r_{BC}^2} (\bm{n}_{BA} \cdot\bm{n}_{BC}) (\bm{v}_{BC}\cdot\bm{n}_{BC}) +\frac{1}{8} \frac{r_{AB}^3}{r_{BC}^2}(\bm{v}_{BC}\cdot\bm{n}_{BA}) +\frac{1}{4} \frac{r_{AB}^2}{r_{BC}}(\bm{n}_{BA} \cdot\bm{n}_{BC})(\bm{v}_B\cdot\bm{n}_{BA})  \notag\\
      &\qquad\qquad+\frac{1}{4} \frac{r_{AB}^2}{r_{BC}}(\bm{v}_B \cdot\bm{n}_{BC}) \bigg)+\frac{\dot{m}_B}{m_B} \left(- \frac{3}{16} \frac{r_{AB}^3}{r_{BC}^2} (\bm{v}_{BC}\cdot\bm{n}_{BA}) +\frac{9}{16} \frac{r_{AB}^3}{ r_{BC}^2} (\bm{n}_{BA}\cdot\bm{n}_{BC})(\bm{v}_{BC}\cdot\bm{n}_{BC})\right)\notag\\
      &\qquad\qquad-\frac{3}{16} \frac{r_{AB}^3}{ r_{BC}} (\bm{n}_{BA}\cdot\bm{n}_{BC})\frac{\dot{m}_B}{m_B} \frac{\dot{m}_C}{m_C}-\frac{1}{16} \frac{r_{AB}^3}{ r_{BC}} (\bm{n}_{BA}\cdot \bm{n}_{BC})\frac{\ddot{m}_C}{m_C}\Bigg]\notag\\
      &\qquad+\frac{\dot{m}_B}{m_B} \bigg[r_{AB}^2\bigg(2(\bm{v}_A\cdot\bm{a}_B)+\frac{5}{4} (\bm{a}_B\cdot\bm{v}_B)+\frac{1}{2} (\bm{a}_B\cdot\bm{n}_{BA}) (\bm{v}_B\cdot\bm{n}_{BA})\bigg)\notag\\
      &\qquad\qquad+r_{AB}\bigg(\frac{3}{8}(\bm{v}_B\cdot\bm{n}_{BA})v_A^2+\frac{3}{2} (\bm{v}_B\cdot\bm{n}_{BA}) v_B^2+2r(\bm{v}_{B}\cdot\bm{n}_{BA})(\bm{v}_A\cdot\bm{v}_B)-\frac{1}{4}  (\bm{v}_B\cdot\bm{n}_{BA})^3\bigg)\bigg]\notag\\
      &\qquad+\frac{\ddot{m}_B}{m_B} \bigg[\frac{1}{4} r_{AB}^3 (\bm{a}_B\cdot\bm{n}_B)+\frac{1}{8} r_{AB}^2 (\bm{v}_B\cdot\bm{n}_{BA})^2+\frac{3}{8}r_{AB}^2v_A^2-\frac{1}{4} r_{AB}^2 v_B^2+2r_{AB}^2(\bm{v}_A\cdot\bm{v}_B)\bigg]\notag\\
      &\qquad+\frac{\dot{m}_B^2}{m_B^2}\bigg[-\frac{1}{16} r_{AB}^3 (\bm{a}_B\cdot\bm{n}_B)+\frac{1}{4} r_{AB}^2 (\bm{v}_B\cdot\bm{n}_{BA})^2+ r_{AB}^2 v_B^2\bigg]\Bigg\}\Bigg\}+\mathcal{O}(c^{-6}).
\end{align} 
There are $\dot{m}$ terms in the Lagrangian, which mean that the interaction between spacetime background and matter caused by MT can be seen in the 1PN order first and exist in the higher orders as well. The Lagrangian is much more simple for 2-Body system like the binary system, which only includes stars A and B. So that we can write $\sum_{B\neq A}=B$, $\sum_{C\neq B}=A$ and so on. The Lagrangian of binary system is shown in Eq.\eqref{Lagrangian}.

Based on the Lagrangian of this N-body system, we can obtain the Lagrangian in the center-of-mass frame in Eq.\eqref{CoML} and the Lagrangian in the spherical coordinate system in Eq.\eqref{SphericalL} that are needed in the subsequent calculations.

\section{Transformation  to the center-of-mass frame}\label{AppD}
In this Appendix we will derive the transformation to the center-of-mass frame, which is used in Eq.\eqref{Lagrangian} and Eq.\eqref{CoML}. We can get the conservative mass-energy dipole moment by following \cite{Gravity_PoissonWill,1976ApJ210764W}
\begin{equation}
    E_A \bm{x}_A+E_B\bm{x}_B=E_{total}\bm{X},
\end{equation}
where $E_A$ and $E_B$ are the energies of  particles $A$ and $B$  at the 1PN order which can be obtained by integrating the energy-momentum tensor $T^{\mu\nu}$ in Eq.\eqref{Tmunu}. $E_A$ is given by \cite{LucBlanchet_2003,AIHPA_1985__43_1_107_0}
\begin{align}\label{Energy}
    E_A=&m_Ac^2+\left\{\frac{1}{2} m_A v_A^2-\frac{G m_A m_B}{2 r}\right\}+\frac{1}{c^2} \left\{\frac{3}{8} m_A v_A^4 +\frac{G m_A m_B}{r} \left(\frac{19}{8} v_A^2-\frac{7}{8} v_B^2-\frac{5}{4} \frac{G m_A}{r}+\frac{7}{4} \frac{G m_B}{r}\right.\right.\notag\\
    &\left.\left.-\frac{1}{8}(\bm{v}_A\cdot\bm{n})^2+\frac{1}{8}(\bm{v}_B\cdot\bm{n})^2-\frac{1}{4}(\bm{v}_A\cdot\bm{n}) (\bm{v}_B\cdot\bm{n})-\frac{7}{4} (\bm{v}_A\cdot\bm{v}_B)+\frac{1}{4}\frac{\dot{m}_B}{m_B}r(\bm{v}_B\cdot\bm{n}) \right)\right\}.\;\;
\end{align}
$E_{total}=E_A+E_B$ is the energy of binary system, $\bm{X}$ is the position vector of the center-of-mass. Since we consider the conservation of total energy without GW radiation and other acceleration, we can set $\bm{X}=0$ to work in the ceter-of-mass frame. Similarly, the equation of conservative momentum is
\begin{equation}
    \frac{\mathrm{d}}{\mathrm{d}t}(E_A\bm{x}_A)+\frac{\mathrm{d}}{\mathrm{d}t}(E_B\bm{x}_B)=E_{total}\frac{\mathrm{d}\bm{X}}{\mathrm{d}t},
\end{equation}
we can also choose coordinate which $\mathrm{d}\bm{X}/\mathrm{d}t=0$ and $\mathrm{d}\bm{x}_A/\mathrm{d}t=\bm{v}_A,\mathrm{d}\bm{x}_B/\mathrm{d}t=\bm{v}_B$. By solving above equations, we can express $\bm{x}_A$ and $\bm{x}_B,$ in terms of $\bm{x}=\bm{x}_A-\bm{x}_B$, $\bm{v}=\bm{v}_A-\bm{v}_B$ and $\bm{a}=\bm{a}_A-\bm{a}_B$ \cite{PhysRevD.63.062005,VanessaCdeAndrade_2001,LucBlanchet_2003}:
\begin{align}
    \bm{x}_A&=+\frac{1}{2} (1-\Delta ) \bm{x}+\mathcal{P}_1\bm{x}+\mathcal{Q}_1\bm{v},\qquad\bm{v}_A=+\frac{1}{2} (1-\Delta )\bm{v}+\mathcal{P}_2\bm{x}+\mathcal{Q}_2\bm{v},\qquad \bm{a}_A=+\frac{m_B}{M}\bm{a},\\
    \bm{x}_B&=-\frac{1}{2} (1+\Delta ) \bm{x}+\mathcal{P}_1\bm{x}+\mathcal{Q}_1\bm{v},\qquad\bm{v}_B=-\frac{1}{2} (1+\Delta )\bm{v}+\mathcal{P}_2\bm{x}+\mathcal{Q}_2\bm{v},\qquad  \bm{a}_B=-\frac{m_A}{M}\bm{a}.
\end{align}
this relationship will be used in the 2PN Lagrangian in Eq.\eqref{Lagrangian} and Eq.\eqref{CoML}. $\mathcal{P}$ and $\mathcal{Q}$ denote the coefficients of $\bm{x}$ and $\bm{v}$ direction individually. Considering MT we have
\begin{align}
    \mathcal{P}_1&=\frac{1}{c^2}\bigg\{\Delta  \eta   \left( \frac{1}{2}v^2-\frac{1}{2}\frac{G M}{r}\right)\bigg\}\notag\\
    &+\frac{1}{c^4}\bigg\{ \bigg[\frac{ G^2 M^2 }{r^2}\Big(\frac{7}{4}-\frac{1}{2}\eta\Big)+\frac{ G M   }{ r}\Big(\frac{3}{4}\eta  \dot{r}^2-\frac{1}{8}\dot{r}^2+\frac{3 }{2}  \eta  v^2+\frac{19}{8}   v^2\Big)-\frac{3}{2}    \eta v^4+\frac{3}{8}   v^4\bigg]\Delta  \eta-\frac{1}{4} G \eta \dot{r} \chi\bigg\},\\
    \mathcal{Q}_1&=\frac{1}{c^4}\bigg\{-\frac{7}{4} G M \Delta  \eta \dot{r}\bigg\},\\
    \mathcal{P}_2&=\frac{1}{c^2} \bigg\{ \frac{G M \Delta  \eta  \dot{r}}{2 r^2}\bigg\}\notag\\
    &+\frac{1}{c^4} \bigg\{\Delta\eta\bigg[\frac{G^2 M^2 }{r^3}\Big(-\frac{3}{2}\eta\dot{r}-\frac{9}{4} \dot{r}\Big)+ \frac{G M }{r^2}\Big(\frac{3}{8}\dot{r}^3-\frac{3}{4}  \eta \dot{r}^3+ \eta v^2 \dot{r}-\frac{9}{8}  v^2 \dot{r}\Big)\bigg]-\frac{G}{4r} \left(G M+r \left(\dot{r}^2-v^2\right)\right)\eta \chi\bigg\},\\
    \mathcal{Q}_2&=\frac{1}{c^2} \bigg\{-\frac{G M \Delta  \eta }{2 r}+\frac{1}{2} \Delta  \eta  v^2\bigg\}\notag\\
    &+\frac{1}{c^4} \bigg\{\Delta  \eta\bigg( \frac{G^2 M^2  }{ r^2}\Big(\frac{7}{2}-\frac{1}{2} \eta \Big)+ \frac{G M }{ r}\Big(\frac{3}{4}\eta  \dot{r}^2+\frac{13}{8} \dot{r}^2+\frac{3}{2} \eta  v^2+\frac{5}{8} v^2\Big)-\frac{3}{2} \eta  v^4+\frac{3}{8}  v^4\bigg)-\frac{1}{4} G \eta \dot{r} \chi\bigg\},
\end{align}
noting that 2PN terms with MT will be ignored as $\mathcal{O}(\chi^2)$ in the expression of acceleration. the MT effect $\chi$ here comes from MT terms in the energy $E_A,E_B$ of the binary star. 

\section{Lagrangian and equation of motion in spherical coordinate frame}\label{AppE}
Based on the Lagrangian in center-of-mass frame in Eq.\eqref{CoML} and the orthogonal frame in Eq.\eqref{orthogonalframe}, we can get the Lagrangian in spherical coordinate frame, which will make us more convenient to calculate the orbital angular momentum, angular acceleration and radial acceleration of the system.
\begin{align}\label{SphericalL}
    \mathcal{L}_S&=\frac{G M^2 \eta}{r}+\frac{1}{2} M \eta \big(\omega^2 r^2+ \dot{r}^2\big)\notag\\
    &+\frac{1}{c^2} \bigg\{-\frac{G^2 M^3 \eta}{2 r^2}+\frac{GM^2\eta}{r} \left(\frac{1}{2} \eta \omega^2 r^2+\frac{3}{2}\omega^2 r^2+ \eta\dot{r}^2+ \frac{3}{2}\dot{r}^2\right)\notag\\
    &\qquad+M\eta\bigg(-\frac{3}{8} \eta \omega^4 r^4+\frac{1}{8} \omega^4 r^4-\frac{3}{4} \eta \omega^2 r^2 \dot{r}^2+\frac{1}{4} \omega^2 r^2 \dot{r}^2-\frac{3}{8} \eta \dot{r}^4+\frac{1}{8} \dot{r}^4\bigg)-\frac{1}{4} G M \Delta \chi \dot{r}\bigg\}\notag\\
    &+\frac{1}{c^4} \bigg\{\frac{G^3M^4\eta}{r^3} \left(\frac{15}{4} \eta+\frac{1}{2}\right)+\frac{G^2M^3\eta}{r^2} \left(\frac{1}{2} \eta^2 \omega^2 r^2-\frac{27}{8} \eta \omega^2 r^2+\frac{7}{4} \omega^2 r^2+2 \eta^2 \dot{r}^2 +\frac{7}{4} \eta \dot{r}^2 +\frac{9}{4} \dot{r}^2\right)\notag\\
    &\qquad+\frac{GM^2\eta}{r} \bigg(-\frac{9}{8} r^4 \eta^2 \omega^4-\frac{3}{8} r^4 \eta \omega^4+\frac{7}{8} r^4 \omega^4-\frac{7}{2} r^2 \eta^2 \dot{r}^2 \omega^2-\frac{27}{4} r^2 \eta \dot{r}^2 \omega^2+\frac{7}{4} r^2 \dot{r}^2 \omega^2+\frac{7}{8} r^3 \eta \ddot{r} \omega^2-\frac{7}{4}  r^2 \eta \dot{r} \dot{\omega} \omega\notag\\
    &\qquad\qquad\qquad\quad+\frac{7 \dot{r}^4}{8 }-\eta r \dot{r}^2 \ddot{r}-2 \eta^2 \dot{r}^4- \eta \dot{r}^4\bigg)\notag\\
    &\qquad+\frac{M\eta}{16}\bigg(\Big(13 \eta^2 -7 \eta + 1\Big)r^6 \omega^6+\Big(39 \eta^2  -21 \eta +3 \Big) r^4  \dot{r}^2 \omega^4+\Big(39 \eta^2 -21 \eta +3\Big)  r^2  \dot{r}^4 \omega^2+\Big(13  \eta^2 -7\eta +1\Big) \dot{r}^6\bigg)\notag\\
    &\qquad+ \bigg[GM\left(-\frac{5}{4} \dot{r}^3+\frac{43}{8} \eta \dot{r}^3-\frac{23}{4}  r^2 \omega^2 \dot{r}+16  r^2 \eta \omega^2 \dot{r}+\Big(\frac{15}{4}  \eta-\frac{7}{4}\Big)r \ddot{r} \dot{r}+\Big(\frac{13}{4} \eta-\frac{5}{4} \Big)r^2 \omega \dot{\omega}\right) \notag\\
    &\qquad\qquad+\frac{G^2 M^2}{r}\left(\frac{ \eta }{2}-\frac{3 }{8}\right)\dot{r}\bigg] \Delta\chi\bigg\}+\mathcal{O}(c^{-6}).
\end{align}
Solving the Euler-Lagrange equation with generalized coordinates $(r,\phi)$ like Eq.\eqref{E-L}, we get accelerations of $r$ and $\phi$
\begin{align}
    \ddot{r}&=-\frac{G M}{r^2}+r \omega^2+\frac{\Delta  \chi}{M \eta}\dot{r}\notag\\
    &+\frac{1}{c^2}\bigg\{\frac{G^2M^2}{r^3}\bigg(2 \eta +4 \bigg) +\frac{GM}{r^2}\left(3 \dot{r}^2-\frac{7}{2} \eta \dot{r}^2-r^2 \omega^2-3 \eta r^2\omega^2\right)+ \left(\frac{3 }{2}\eta\dot{r}^2-\dot{r}^2-\frac{3}{2}\eta r^2 \omega^2 +r^2 \omega^2+\frac{2 GM}{r}\eta \right) \frac{\Delta  \chi}{M \eta}\dot{r}\bigg\}\notag\\
    &+\frac{1}{c^4}\bigg\{\frac{G^3M^3}{r^4} \left(-9-\frac{87 \eta }{4}\right)+\frac{G^2M^2}{r^3} \left(-5 \eta ^2 r^2\omega^2- \frac{1}{2}\eta  r^2\omega^2+11 \eta  \dot{r}^2-4 \eta ^2 \dot{r}^2\right)\notag\\
    &\qquad+\frac{GM}{r^2} \left(\frac{5}{2} r^4 \omega^4+\frac{11}{2} \eta ^2r^4  \omega^4-3 \eta  r^4\omega^4+\frac{71}{2} \eta ^2 \dot{r}^2r^2 \omega^2-5 \dot{r}^2 r^2\omega^2-\frac{35}{2} \eta  \dot{r}^2 r^2\omega^2+\frac{21}{8} \eta ^2 \dot{r}^4+\frac{21}{8}\eta \dot{r}^4\right)\notag\\
    &\qquad+\frac{1}{2} r^5 \omega^6+\frac{9}{2} r^5 \eta ^2 \omega^6-3 r^5 \eta  \omega^6-r^3 \dot{r}^2 \omega^4-9 r^3 \eta ^2 \dot{r}^2 \omega^4+6 r^3 \eta  \dot{r}^2 \omega^4+\frac{81}{2} r \eta ^2 \dot{r}^4 \omega^2+\frac{9}{2} r \dot{r}^4 \omega^2-27 r \eta  \dot{r}^4 \omega^2\notag\\
    &\qquad+\bigg[\frac{G^2M^2}{r^2} \left(4 \eta^2  +10\eta-\frac{9}{4}\right)+\frac{GM}{r} \left(6 \eta^2  \dot{r}^2-\frac{3}{2} \eta\dot{r}^2-4 \dot{r}^2-3 \eta^2 r^2  \omega^2 -7 \eta r^2 \omega^2+\frac{19}{4} r^2 \omega^2 \dot{r}\right)\notag\\
    &\qquad\qquad-3 \eta^2  \dot{r}^4+\frac{9}{8} \eta\dot{r}^4+\frac{81}{4} \eta r^2 \omega^2 \dot{r}^2-24 \eta^2 r^2   \omega^2 \dot{r}^2-4 r^2 \omega^2 \dot{r}^2+\frac{81}{8} r^4 \omega^4-12  \eta^2 r^4 \omega^4 -2 r^4 \omega^4 \bigg] \frac{\Delta \chi}{M\eta}\dot{r}\bigg\}\nonumber\\
    &+\mathcal{O}(c^{-6}),
    \end{align}
    \begin{align}
    \frac{\dot{\omega}}{\omega}&=-\frac{2 \dot{r}}{r}+\frac{\Delta \chi}{M \eta}\notag\\
    &+\frac{1}{c^2}\bigg\{\bigg(2 r^2 \omega^2-6  \eta r^2\omega^2+\frac{GM}{r} \Big(4 \eta +2 \Big)\bigg)\frac{\dot{r}}{r}+ \left(\frac{3}{2} \eta r^2 \omega^2-r^2 \omega^2+\dot{r}^2-\frac{3}{2}\eta \dot{r}^2+\frac{G M\eta}{r}\right) \frac{\Delta \chi}{M \eta}\bigg\}\notag\\
    &+\frac{1}{c^4}\bigg\{\frac{G^2M^2}{r^2} \left(-4 \eta ^2 -\frac{73}{2} \eta-2\right)\frac{\dot{r}}{r}+\frac{GM}{r} \left(26 \eta ^2 r^2\omega^2 -\frac{17}{2} \eta  r^2 \omega^2+26 \eta ^2 \dot{r}^2-20 \eta  \dot{r}^2+8 r^2\omega^2 +2 \dot{r}^2\right)\frac{\dot{r}}{r}\notag\\
    &\qquad+\bigg(-18 \eta ^2 \omega^4 r^4 +12 \eta  \omega^4 r^4-18 \eta ^2 \omega^2 r^2 \dot{r}^2+12 \eta  \omega^2 r^2 \dot{r}^2-2 \omega^4 r^4-2 \omega^2 r^2 \dot{r}^2\bigg)\frac{\dot{r}}{r}\notag\\
    &\qquad+ \bigg[\frac{G^2M^2}{r^2} \left(\eta^2 -\frac{29}{4}\eta-\frac{3}{4}\right)+\frac{GM}{r} \left(3 \eta^2 r^2 \omega^2 -\frac{13}{4} r^2\omega^2 +4 \dot{r}^2-6 \eta^2  \dot{r}^2+2\eta \dot{r}^2-6\dot{r}^2 \right)\notag\\
    &\qquad\qquad-3 \eta^2  \omega^4 r^4+\frac{9}{8}\eta \omega^4 r^4-24 \eta^2  \omega^2 r^2 \dot{r}^2-4 \omega^2 r^2 \dot{r}^2+\frac{81}{4}\eta \omega^2 r^2 \dot{r}^2-12 \eta^2  \dot{r}^4-2 \dot{r}^4+\frac{81}{8} \eta \dot{r}^4\bigg]\frac{\Delta \chi}{M \eta}\bigg\}\nonumber\\
    &+\mathcal{O}(c^{-6}),\label{domega}
\end{align}

Let's come back to the Kepler third law, in the orthogonal frame we can get the expression of angular velocity $\omega$ with $GM/r$ by solving the equation of acceleration in $\bm{n}$ direction as:
\begin{align}
    \omega^2&=\frac{G M}{r^3}+\frac{1}{c^2} \bigg\{\frac{G^2M^2}{r^4} \left( \eta-3\right)+\frac{GM}{r^3} \left(\frac{7}{2} \eta \dot{r}^2-3 \dot{r}^2\right)-\frac{G \Delta \chi \dot{r}}{r^2}\bigg\}\notag\\
    &+\frac{1}{c^4} \bigg\{\frac{G^3M^3}{r^5} \left( \eta^2+\frac{41}{4}\eta+6\right)+\frac{G^2M^2}{r^4} \left(\frac{21}{2} \eta^2 \dot{r}^2-\frac{45}{2 } \eta \dot{r}^2-3 \dot{r}^2\right)+\frac{GM}{r^3} \left(-\frac{21}{8} \eta^2 \dot{r}^4-\frac{21}{8} \eta \dot{r}^4\right)\notag\\
    &\qquad+ \left(\frac{G^2M^2}{r^3} \left(-7 \eta^2-\frac{49}{8} \eta+\frac{11}{2} \right)+\frac{GM}{r^2} \left(4 \dot{r}^2-\frac{3}{4} \dot{r}^2\right)+\frac{3 \eta^2 \dot{r}^4}{ r}-\frac{9 \eta\dot{r}^4}{8 r}\right)\frac{\Delta \chi}{M\eta}\dot{r}\bigg\}+\mathcal{O}(c^{-6}).
\end{align}
At the leading order we can get the relationship between $\dot{r}$ and $\dot{\omega}$ by taking the derivative of both sides of the equation with respect to time, putting it back to Eq.\eqref{domega} and getting
\begin{equation}
    \frac{\dot{r}}{r}=-\frac{2}{3}\frac{\dot{\omega}}{\omega},\qquad \frac{\dot{\omega}}{\omega}=-3\frac{\Delta\chi}{M\eta},
\end{equation}
this effect will appear at the 0PN order. It should be noted that even in the quasi-circular orbit, we cannot directly remove the terms of $\dot{r}$ , because in addition to the GW radiation effect of 2.5PN, the mass transfer begins to have an effect at 0PN, and we need to further analyze the magnitude of this effect. Similarly, we can use this relationship to derive out the radius $r$ at 1PN order
\begin{align}
    r_{1PN}&={(G M)}^{1/3}\omega^{-2/3}+\frac{1}{c^2} \bigg\{G M\left(\frac{1}{3}  \eta -1\right)+\frac{\Delta^2\chi^2}{M^2\eta^2}  GM\left(4 \eta-4 \right)\omega^{-2}\bigg\}\notag\\
    &={(G M)}^{1/3}\omega^{-2/3}+\frac{1}{c^2} \bigg\{G M\left(\frac{1}{3}  \eta -1\right)+\mathcal{O}(\chi^2)\bigg\}.
\end{align}
We can find that this equation include terms with MT in $\chi^2$ order and without $\chi$ order, so we can treat $\mathcal{O}(\chi^2)$ as a small quantity. By iterating the results of 1PN into the calculations of 2PN, we can obtain
\begin{align}
    r={(G M)}^{1/3}\omega^{-2/3}+\frac{1}{c^2} G M\left(\frac{1}{3}  \eta -1\right)+\frac{1}{c^4}  (G M)^{5/3}\omega^{2/3} \left(\frac{1}{9}  \eta^2+\frac{19}{4}\eta\right)+\mathcal{O}(\chi^2)+\mathcal{O}(c^{-6}).
\end{align}
Taking the derivative of the above equation with respect to time and then dividing by r, we can obtain. 
\begin{align}\label{romega}
    \frac{\dot{r}}{r}=&-\frac{2}{3} \frac{\dot{\omega}}{\omega}+\frac{1}{c^2} \bigg\{\frac{\dot{\omega}}{\omega} \left(\frac{2}{9}  \eta-\frac{2}{3} \right)-\frac{\Delta \chi}{3 M}\bigg\}(G M)^{2/3}\omega^{2/3}\notag\\
    &+\frac{1}{c^4} \bigg\{\frac{ \dot{\omega}}{\omega} \left(\frac{2}{27}  \eta^2+\frac{61}{9} \eta-\frac{2}{3}\right)+\frac{\Delta \chi}{M} \left(-\frac{61}{12}-\frac{1}{9}\eta\right)\bigg\}(G M)^{4/3}\omega^{4/3}+\mathcal{O}(c^{-6}),
\end{align}
then we can put this results into Eq.\eqref{domega}, replace $\dot{r}$ and $r$ by $\chi$ and $\dot{\omega}$, then we can get the variation of orbit angular velocity in Eq.\eqref{variationoemga}. So far, we have found that the expression of the distance $r$ of the quasi-circular orbit binary stars will not change when the MT effect $\chi$ is retained to the first order, but the derivative of the distance with time $\dot{r}$ will be affected, and this effect will also be substituted into the expression of the acceleration of angular velocity $\dot{\omega}$. If we set $\chi=0$, all of above results will comeback to the normal form calculated by Blanchet \cite{Blanchet2014}
.

\bibliography{BIB.bib}
\bibliographystyle{unsrt}

\end{document}